\documentclass[runningheads]{llncs}

\usepackage{graphicx}

\usepackage{tikz}
\usepackage{comment}
\usepackage{amsmath,amssymb} 
\usepackage{color}

\usepackage{graphicx}
\usepackage{cases}
\usepackage{setspace}
\usepackage{overpic}
\usepackage{rotating}
\graphicspath{{figures/}}

\usepackage{subeqnarray}
\usepackage{multirow}
\usepackage{tabularx}
\usepackage{epstopdf}
\usepackage{subfigure}
\usepackage{enumerate}
\usepackage{xcolor}
\usepackage[pagebackref=true,breaklinks=true,letterpaper=true,colorlinks=true,bookmarks=false]{hyperref}

\def\ie{\emph{i.e.,\ }}

\begin{document}
\pagestyle{headings}
\mainmatter
\def\ECCVSubNumber{100}  

\title{AIM 2020 Challenge on Efficient
Super-Resolution: Methods and Results} 

\titlerunning{AIM 2020 Challenge on Efficient Super-Resolution}
%
\author{Kai Zhang$^*$ \and
Martin Danelljan$^*$ \and
Yawei Li$^*$ \and
Radu Timofte$^*$ \and
Jie Liu \and
Jie Tang \and
Gangshan Wu \and
Yu Zhu \and
Xiangyu He \and
Wenjie Xu \and
Chenghua Li \and
Cong Leng \and
Jian Cheng \and
Guangyang Wu \and
Wenyi Wang \and
Xiaohong Liu \and
Hengyuan Zhao \and
Xiangtao Kong \and
Jingwen He \and
Yu Qiao \and
Chao Dong \and
Xiaotong Luo \and
Liang Chen \and
Jiangtao Zhang \and
Maitreya Suin \and
Kuldeep Purohit \and
A. N. Rajagopalan \and
Xiaochuan Li \and
Zhiqiang Lang \and
Jiangtao Nie \and
Wei Wei \and
Lei Zhang \and
Abdul Muqeet \and
Jiwon Hwang \and
Subin Yang \and
JungHeum Kang \and
Sung-Ho Bae \and
Yongwoo Kim \and
Liang Chen \and
Jiangtao Zhang \and
Xiaotong Luo \and
Yanyun Qu \and
Geun-Woo Jeon \and
Jun-Ho Choi \and
Jun-Hyuk Kim \and
Jong-Seok Lee \and
Steven Marty \and
Eric Marty \and
Dongliang Xiong \and
Siang Chen \and
Lin Zha \and
Jiande Jiang \and
Xinbo Gao \and
Wen Lu \and
Haicheng Wang \and
Vineeth Bhaskara \and
Alex Levinshtein \and
Stavros Tsogkas \and
Allan Jepson \and
Xiangzhen Kong \and
Tongtong Zhao \and
Shanshan Zhao \and
Hrishikesh P S \and
Densen Puthussery \and
Jiji C V \and
Nan Nan \and
Shuai Liu \and
Jie Cai \and
Zibo Meng \and
Jiaming Ding \and
Chiu Man Ho \and
Xuehui Wang \and
Qiong Yan \and
Yuzhi Zhao \and
Long Chen \and
Jiangtao Zhang \and
Xiaotong Luo \and
Liang Chen \and
Yanyun Qu \and
Long Sun \and
Wenhao Wang \and
Zhenbing Liu \and
Rushi Lan \and
Rao Muhammad Umer \and
Christian Micheloni
}
%
\authorrunning{K. Zhang et al.}
%
\institute{}
\maketitle

\let\thefootnote\relax\footnotetext{$^*$ K. Zhang (kai.zhang@vision.ee.ethz.ch, Computer Vision Lab, ETH Zurich), M. Danelljan, Y. Li and R. Timofte were the challenge organizers, while the other authors participated in the challenge. Appendix~\ref{sec:teams} contains the authors' teams and affiliations.
AIM webpage: \url{https://data.vision.ee.ethz.ch/cvl/aim20/}}

\begin{abstract}
This paper reviews the AIM 2020 challenge on efficient single image super-resolution with focus on the proposed solutions and results.
The challenge task was to super-resolve an input image with a magnification factor $\times$4 based on a set of prior examples of low and corresponding high resolution images.
The goal is to devise a network that reduces one or several aspects such as runtime, parameter count, FLOPs, activations, and memory consumption while at least maintaining PSNR of MSRResNet. The track had 150 registered participants, and 25 teams submitted the final results. They gauge the state-of-the-art in efficient single image super-resolution.

\end{abstract}

\section{Introduction}
Single image super-resolution (SR) aims at recovering a high-resolution (HR) image from a single degraded low-resolution (LR) image. Since the dawn of deep learning, this problem has been frequently tackled by researchers from low-level vision community with models based on convolutional neural networks (CNN)~\cite{dong2014learning,Kim-CVPR-2016,shi2016real,ledig2017photo}. 
In order to strive towards the ultimate goal of deploying SR models for real-world applications, there exist a number of important research directions. The most popular direction is to improve PSNR or the perceptual quality, based on bicubic degradation assumption~\cite{Timofte-ACCV-2014,zhang2020ntire}. Significant achievements have been made on designing network architectures and training losses for this purpose~\cite{ledig2017photo,wang2018esrgan}. However, such bicubic degradation based methods would give rise to poor performance if the real degradation deviates from the assumed one~\cite{efrat2013accurate}. Hence, another direction is to design a network to handle a more general degradation with varying factors such as blur kernel~\cite{zhang2017learning,zhang2018learning,zhang2020deep,zhang2020plug}. In practical applications, the blur kernel is usually unknown, thus some researchers attempt to estimate the blur kernel of a given LR image for better reconstruction~\cite{bell2019blind}. Such a strategy has also been successfully applied to the direction of SR with unpaired data~\cite{lugmayr2019unsupervised,lugmayr2020ntire}, where even more general degradation operations are considered. A recently emerging direction is to account for the ill-posed nature of the SR problem by learning stochastic and explorable LR to HR mappings using GAN~\cite{bahat2019explorableSR,menon2020pulse} or Normalizing Flow based~\cite{SRFlow} approaches.
This challenge report focuses on another research direction, namely that of \emph{efficient SR}, which is of crucial importance in order to deploy models on resource-constrained devices.

There are many factors that affect the efficiency of an SR network. Some typical factors are runtime, the number of parameters, and floating point operations (FLOPs).
During past few years, several efficient SR works have been proposed based on different techniques, including hand-designed network architectures~\cite{dong2014learning,Kim-CVPR-2016,zhang2018residual,hui2019lightweight}, network pruning~\cite{li2020dhp}, filter decomposition~\cite{li2019learning}, network quantization~\cite{liu2018bi,li2019additive}, neural architecture search (NAS)~\cite{liu2018darts,cai2019once}, and knowledge distillation~\cite{hinton2015distilling,yin2020dreaming}. Despite of significant achievements, these methods mostly focus on the number of parameters and FLOPs. Recent works on high-level tasks have pointed out that fewer FLOPs does not always indicate better network efficiency, and the number of network activations is instead a more accurate measure of the network efficiency~\cite{radosavovic2020designing}. As a result, efficient SR methods require a thorough analysis from different aspects rather than only from the aspects of parameters and FLOPs.

Jointly with the Advances in Image Manipulation (AIM) 2020 workshop, we organize the AIM Challenge on Efficient Super-Resolution. The task of the challenge is to super-resolve an LR image to an HR image with a magnification factor $\times$4 by a network that reduces one or several aspects such as runtime, parameters, FLOPs, activations and memory consumption, while at least maintaining PSNR of the baseline model. The challenge aims to seek advanced and novel solutions for efficient SR, to benchmark their efficiency, and identify the general trends.

\section{AIM 2020 Efficient Super-Resolution Challenge}
This challenge is one of the AIM 2020 associated challenges on:
scene relighting and illumination estimation~\cite{elhelou2020aim_relighting}, image extreme inpainting~\cite{ntavelis2020aim_inpainting}, learned image signal processing pipeline~\cite{ignatov2020aim_ISP}, rendering realistic bokeh~\cite{ignatov2020aim_bokeh}, real image super-resolution~\cite{wei2020aim_realSR}, efficient super-resolution~\cite{zhang2020aim_efficientSR}, video temporal super-resolution~\cite{son2020aim_VTSR} and video extreme super-resolution~\cite{fuoli2020aim_VXSR}.
The objectives of this challenge are:
(i) to advance research on efficient SR; (ii) to compare the efficiency of different methods and (iii) to offer an opportunity for academic and industrial attendees to interact and explore collaborations. This section details the challenge itself.

\subsection{DIV2K Dataset~\cite{Agustsson_2017_CVPR_Workshops}}
Following~\cite{Agustsson_2017_CVPR_Workshops}, the DIV2K dataset is adopted, which contains 1,000 DIVerse 2K resolution RGB images. The HR DIV2K is divided into 800 training images, 100 validation images and 100 testing images.
The corresponding LR DIV2K in this challenge is the bicubicly downsampled counterpart with a down-scaling factor $\times 4$.
The testing HR images are hidden from the participants during the whole challenge.

\subsection{MSRResNet Baseline Model}
\label{sec:msrresnet}
The MSRResNet~\cite{wang2018esrgan} serves as the reference SR model in this challenge. The aim is to improve its efficiency while maintaining the SR performance. The MSRResNet contains 16 residual blocks and a global identity skip connection is adopted. Specifically, each residual block of MSRResNet consists of two $3\times3$ convolutional layers with Leaky ReLU activation in the middle and an identity skip connection summed to its output,
while the global identity skip connection directly sums the bilinearly interpolated LR image to the output of final convolutional layer. The reference MSRResNet is trained on DIV2K~\cite{Agustsson_2017_CVPR_Workshops}, Flickr2K and OST~\cite{wang2018recovering} datasets. The quantitative performance and efficiency metrics  of MSRResNet are given as follows. (1) The number of parameters is 1,517,571 (1.5M). (2) The average PSNRs on validation and testing sets of DIV2K are 29.00 dB and 28.70dB, respectively. (3) The average runtime over validation set with PyTorch 1.5.1, CUDA Toolkit 10.2, cuDNN 7.6.2 and a single Titan Xp GPU is 0.110 seconds. (4) The number of FLOPs for an input of size $256\times256$ is 166.36G. (5) The number of activations (\ie elements  of  all  outputs of convolutional layers) for an input of size $256\times256$ is 292.55M. (5) The maximum GPU memory consumption for an input of size $256\times256$ is 610M. (6) The number of convolutional layers is 37.

\subsection{Competition}
The aim of this challenge is to devise a network that reduces one or several aspects such as runtime, parameters, FLOPs, activations and memory consumption while at least maintaining the PSNR of MSRResNet.

\noindent{\textbf{Challenge phases }}
\textit{(1) Development and validation phase:} The participants had access to the 800 LR/HR training image pairs and 100 LR validation images of the DIV2K dataset. The participants were also provided the reference MSRResNet model from github (\url{https://github.com/znsc/MSRResNet}), allowing them to benchmark its runtime on their system, and to adopt it as a baseline if desired. 
The participants could upload the HR validation results on the evaluation server to measure the PSNR of their model to get immediate feedback. The number of parameters and runtime was computed by the participant.
\textit{(2) Testing phase:} 
In the final test phase, the participants recieved access to the 100 LR testing images. The participants then submitted their super-resolved results to the Codalab evaluation server and e-mailed the code and factsheet to the organizers.
The organizers verified and ran the provided code to obtain the final results. Finally, the participants received the final results at the end of the challenge.

\noindent{\textbf{Evaluation protocol }}
The quantitative evaluation metrics includes validation and testing PSNRs, runtime, number of parameters, number of FLOPs, number of activations, and maximum GPU memory consumed during inference. The PSNR was measured by first discarding the 4-pixel boundary around the images.
The runtime is averaged over the 100 LR validation images and the best one among three consecutive trails is selected as the final result. The FLOPs, activations, and memory consumption are evaluated on an input image of size $256\times256$.
Among the above metrics, the runtime is regarded as the most important one. The validation and testing PSNRs should be at least on par with the baseline. 
A code example for calculating these metrics is available at \url{https://github.com/cszn/KAIR/blob/master/main_challenge_sr.py}.

\begin{table*}[!t]
\caption{Results of AIM 2020 efficient SR challenge. `*' means the organizers did not verify the results. `Runtime' is tested on validation datasets, the average image size is 421$\times$421. `\#Params' denotes the total number of parameters. `FLOPs' is the abbreviation for floating point operations. `\#Activations' measures the number of elements of all outputs of convolutional layers. `Memory' represents maximum GPU memory consumption according to the PyTorch function $\texttt{torch.cuda.max\_memory\_allocated()}$. `\#Conv' represents the number of convolutional layers. `FLOPs', `\#Activations', and `Memory' are tested on an LR image of size 256$\times$256. \textbf{This is not a challenge for PSNR improvement. The `validation/testing PSNR' and `\#Conv' are not ranked}.
}
\centering
\resizebox{\linewidth}{!}
{
\begin{tabular}{ll||ll|lllll|ccccc<{\centering}}
 \multirow{2}{*}{Team} &  \multirow{2}{*}{Author}  &   PSNR&   PSNR&   Runtime&\#Params  & \#FLOPs & \#Activations  & Memory &  \multirow{2}{*}{\#Conv} &  Extra \\

    &   & [Val.]  & [Test] & [Val.] [s] & [M] &   [G] & [M]  & [M] &  &  Data      \\\hline\hline

NJU\_MCG &TinyJie & 29.04 & 28.75  & 0.037$_{(1)}$  & 0.433$_{(3)}$ & 27.10$_{(2)}$ & 112.03$_{(1)}$ & 200$_{(4)}$ & 64 & Yes  \\

AiriA\_CG&Now& 29.00 & 28.70  & 0.037$_{(1)}$  & 0.687$_{(10)}$ & 44.98$_{(9)}$ & 118.49$_{(2)}$ & 168$_{(3)}$ & 33 & Yes  \\

UESTC-MediaLab &Mulns &  29.01 & 28.70  & 0.060$_{(4)}$  & 0.461$_{(5)}$ & 30.06$_{(4)}$ & 219.61$_{(10)}$ & 146$_{(2)}$ & 57 & Yes \\

XPixel &zzzhy & 29.01  & 28.70  &  0.066$_{(6)}$ & 0.272$_{(1)}$ & 32.19$_{(6)}$& 270.53$_{(12)}$ & 311$_{(8)}$  & 121  &  Yes\\

HaiYun &Sudo & 29.09  & 28.78  & 0.058$_{(3)}$  & 0.777$_{(13)}$ & 49.67$_{(10)}$ & 132.31$_{(4)}$ & 225$_{(5)}$ & 104  &  Yes \\

IPCV\_IITM & ms\_ipcv & 29.10  & 28.68  & 0.064$_{(5)}$  & 0.761$_{(12)}$ & 50.85$_{(11)}$& 130.41$_{(3)}$  & 229$_{(6)}$  &  59 &  Yes \\

404NotFound&xiaochuanLi& 29.01 & 28.70  & 0.073$_{(9)}$  & 0.599$_{(8)}$ & 39.36$_{(7)}$ & 170.06$_{(6)}$ & 271$_{(7)}$ & 90 & Yes  \\

MDISL-lab&ppplang & 29.01 & 28.68 &  0.067$_{(7)}$ & 0.660$_{(9)}$ & 42.40$_{(8)}$ & 149.09$_{(5)}$ & 516$_{(12)}$ & 61 & Yes  \\

MLVC &ysb & 29.00 & 28.72  & 0.104$_{(11)}$  & 0.441$_{(4)}$ & 27.11$_{(3)}$ & 212.24$_{(9)}$ & 112$_{(1)}$ & 159 & No  \\

XMUlab &SuckChen & 29.00  & 28.77  & 0.078$_{(10)}$  & 0.691$_{(11)}$ & 53.62$_{(12)}$  & 184.74$_{(7)}$ & 468$_{(10)}$ & 72 & No  \\

MCML-Yonsei&GWJ & 29.01 & 28.66  &  0.070$_{(8)}$ & 1.289$_{(15)}$ & 84.43$_{(14)}$ & 188.74$_{(8)}$ & 798$_{(16)}$ & 68 & Yes  \\

LMSR&martyste & 29.00 & 28.71  & 0.081$_{(11)}^*$  & 1.126$_{(14)}^*$ & 75.72$_{(14)}^*$ & 158.33$_{(6)}^*$ & 192$_{(4)}^*$ & 31* & Yes  \\

ZJUESR2020 &BearMaxZJU & 29.04  & 28.74  & 0.105$_{(12)}$  & 0.516$_{(6)}$ & 54.38$_{(13)}$ & 225.44$_{(11)}$ &  594$_{(13)}$ & 42  & Yes \\

SC-CVLAB&chensa& 29.01 & 28.72  & 0.157 & 0.353$_{(2)}$ & 26.96$_{(1)}$ & 302.30$_{(13)}$ & 595$_{(15)}$ & 91 & Yes  \\

HiImageTeam &HiImageTeam & 29.01  & 28.68  & 0.153  & 0.530$_{(7)}$ & 90.11$_{(15)}$ &325.05$_{(14)}$ & 378$_{(9)}$  & 101 &  Yes \\

SAMSUNG\_TOR\_AIC &hcwang &  28.98 & 28.71  & 0.240  & 0.558$_{(8)}$ & 31.88$_{(5)}$ & 576.45$_{(16)}$ & 477$_{(11)}$ & 59 & Yes  \\

neptuneai&neptuneai& 29.14 & 28.84  & 0.217  & 1.227$_{(14)}$ & 147.72$_{(16)}$ & 535.82$_{(15)}$ & 597$_{(15)}$ & 45 & *   \\

lyl&tongtong & 29.44 & 29.13  & *  & 0.408* & * & * & * & 128* & No  \\

CET\_CVLab&hrishikeshps& 29.00 & 28.74  &  5.00 & 1.378* & * & * & * & * & Yes  \\

wozhu &wozhu &  28.98 &  * & *  & 0.720* & * & * & * & * & Yes  \\

\hline
\multicolumn{11}{c}{The following 5 methods are not ranked since their validation/testing PSNRs are not on par with the baseline.}\\
\hline
InnoPeak\_SR&qiuzhangTiTi& 28.93 & 28.60  & 0.053  & 0.361 & 81.72 &145.75 & 66 & 35 & Yes  \\

Summer &sysu\_wxh &  28.87 & 28.54  & 0.043  & 0.488 & 31.82 & 125.30 & 227 & 35 & No  \\

Zhang9678 &Zhang9678 & 28.78  & 28.50  & *  & 0.664* &  48.08* & * & * & 36*  & No  \\

H-ZnCa&suen& 28.69 & 28.42  &  0.045 & 0.364 & 32.01 & 170.45 & 299 & 67 & No  \\

MLP\_SR&raoumer & 27.89 & 27.77  & 1.313  & 0.047 & 50.66* & 351.27* & 1064 & 10* & Yes \\

\hline

\textit{Winner AIM19} & \textit{IMDN}& 29.13 & 28.78 & 0.050& 0.893 & 58.53 & 154.14 & 120 & 43 &  Yes \\

\textit{Baseline} & \textit{MSRResNet} & 29.00 & 28.70 & 0.114 &1.517 & 166.36& 292.55 & 610 & 37 & Yes 

\end{tabular}
}
\label{table_track1}
\end{table*}

\section{Challenge Results}
Table~\ref{table_track1} reports the final test results and rankings of the teams. The solutions with lower validation PSNR than the MSRResNet baseline are not ranked. In addition, the solutions by lyl, LMSR, CET\_CVLab and wozhu teams are not ranked due to the lack of experimental verification by the organizers. 
The results of the overall first place winner team in AIM 2019 constrained SR challenge~\cite{zhang2019aim} are also reported for comparison. The methods evaluated in Table~\ref{table_track1} are briefly described in Sec.~\ref{sec:methods_and_teams} and the team members are listed in Appendix~\ref{sec:teams}.

According to Table~\ref{table_track1}, we can have the following observations. First, the NJU\_MCG team is the
overall first place winner of this challenge, while AiriA\_CG and UESTC-MediaLab win the overall second place and overall third place, respectively.
Second, NJU\_MCG and AiriA\_CG produce the best runtime; XPixel is the first place winner for the number of parameters; SC-CVLab, NJU\_MCG and MLVC are the top-3 teams that achieve similar performance on FLOPs; NJU\_MCG and AiriA\_CG are the first two place winners for the number of activations; MLVC achieves the best performance for memory consumption.
Third, MLVC and SC-CVLAB are superior in the number of parameters and the number of FLOPs but fail to get a matched runtime. On the other hand, although the methods proposed by 404NotFound and MLVC have lower parameters and FLOPs than IMDN, they exhibit a much slower runtime.
To analyze such discrepancies, we report the Spearman  rank-order  correlation  coefficient  (SROCC)  values of the number of parameters, the number of FLOPs, the number of activations, and maximum GPU memory consumption with respect to runtime in Table~\ref{table_rank}. Note that SROCC is widely used to measure the prediction monotonicity of a metric and a better metric tends to have a higher SROCC. It can be seen from Table~\ref{table_rank} that the number of parameters and the number of FLOPs do not correlate well with the runtime. Instead, the number of activations is a better metric. Such a phenomenon has also been reported in~\cite{radosavovic2020designing}.  
Note that the number of parameters and the number of FLOPs
are still important aspects of model efficiency.

\begin{table*}[!t]
\caption{Spearman rank-order correlation coefficient (SROCC) values 
of  \#Params, \#FLOPs, \#Activations, Memory with respect to runtime.}
\centering
\resizebox{0.7\linewidth}{!}
{
\begin{tabular}{c|c|c|c|cc<{\centering}}

~~Metric~~ & ~~\#Params~~  & ~~\#FLOPs~~ & ~~\#Activations~~  & ~~Memory~~  \\\hline\hline
SROCC & 0.1734  & 0.2397 & 0.8737 & 0.6671 \\
\end{tabular}
}
\label{table_rank}
\end{table*}

\subsection{Architectures and main ideas}
Various techniques are proposed to improve the efficiency of MSRResNet and IMDN. Some typical techniques are given in the following.
\begin{enumerate}
\item \textbf{Modifying the information multi-distillation block of IMDN}. The overall first place winner NJU\_MCG proposed an efficient residual feature distillation block (RFDB) by incorporating shallow residual connection and enhanced spatial attention (ESA) module, using $1\times1$ convolutions for feature distillation, and reducing the number of channels from 64 to 50.
AiRiA\_CG proposed to reduce IMDB blocks and adopt converted asymmetric convolution to improve the efficiency.
Inspired by IMDB and IdleBlock, ZJUESR2020 proposed multi-scale IdleBlock.
\item \textbf{Changing the upsampling block}. Instead of achieve a upscaling factor of 4 via two successive `PixelShuffle($\times$2)$\rightarrow$Conv$\rightarrow$Leaky ReLU' as in MSRResNet, XPixel proposed to replace the PixelShuffle layer with nearest neighbor interpolation layer, while most of the other methods, such as NJU\_MCG, AiRiA\_CG, HaiYun and IPCV\_IITM, proposed to directly reconstruct the HR image via a single PixelShuffle($\times$4) layer.
\item \textbf{Adopting global feature aggregation}. In contrast to the local feature aggregation strategy of IMDN, a global feature aggregation strategy which concatenates the features of different blocks is adopted in several teams such as NJU\_MCG, Haiyun, IPCV\_IITM and 404NotFound. As a typical example, NJU\_MCG proposed to concatenate the outputs of 4 RFDB blocks, then use a $1\times1$ convolutional layer for feature reduction and finally adopt `Conv3$\times$3$\rightarrow$PixelShuffle($\times4$)' to produce the HR image.
\item \textbf{Incorporating attention module}. NJU\_MCG proposed to insert enhanced spatial attention module into the RFDB block. Xpixel proposed pixel attention to produce 3D attention maps. MLVC proposed multi-attention block based on enhanced spatial attention (ESA) and cost-efficient attention (CEA).
\item \textbf{Reducing the number of parameters by recursive layers}. Zhang9678 proposed to adopt LSTM to reduce parameters, while InnoPeak\_SR proposed recursive residual blocks.
\item \textbf{Applying network pruning}. SC-CVLAB proposed a fine-grained channel pruning strategy to get a lightweight model from an over-parameterized hybrid composition based SR network.
\item \textbf{Replacing the basic residual block of MSRResNet with new block}. Xpixel proposed self-calibrated convolution block with pixel attention. 404NotFound proposed to replace the normal $3\times3$ convolution with Ghost convolution and $1\times3$ convolution. SAMSUNG\_TOR\_AIC proposed modified MobileNetV3 block.
    
\end{enumerate}

\subsection{Fairness}
There are some fair and unfair tricks to improve the validation and testing PSNRs for this challenge. On one hand, using additional training data is fair since the MSRResNet baseline was trained on DIV2K~\cite{Agustsson_2017_CVPR_Workshops}, Flickr2K~\cite{Timofte_2017_CVPR_Workshops} and OST~\cite{wang2018recovering} datasets. Most of the teams used the provided  DIV2K and additional Flickr2K for training. In addition, using advanced data augmentation strategy during training is also a fair trick. 
On the other hand, it is unfair to train the model with the validation LR images, validation HR images, and testing LR images. 
First, training on LR/HR validation images would improve the validation PSNR.
Second, it tends to get a PSNR gain if the model is trained on pairs of LR images and their downsampled counterparts. 
Third, the PSNR can be improved by knowledge distillation technique on validation and testing LR images.

\subsection{Conclusions}
From the above analysis of different solutions, we can have several conclusions.
(i) The proposed methods improve the state-of-the-art for efficient SR. Compared to the first place method IMDN in AIM 2019 constrained SR challenge, NJU\_MCG team's method provides a significant gain with respect to the runtime, parameters, FLOPs, and activations.
(ii) The number of FLOPs and the number of parameters do not correlate well with network efficiency. In comparison, the number of activations is a more proper metric.
(iii) All of the overall top-6 methods employ hand-designed network architecture. The effectiveness of network pruning, knowledge distillation, network quantization and NAS for this efficient SR challenge requires further study.
(iv) Future work on efficient SR should take runtime,  parameters, FLOPs, and activations into consideration.

\section{Challenge Methods and Teams}
\label{sec:methods_and_teams}
\subsection*{NJU\_MCG}
\begin{figure}[!ht]
    \centering
    \subfigure[]{\label{fig:imdb}\includegraphics[width=.3\linewidth]{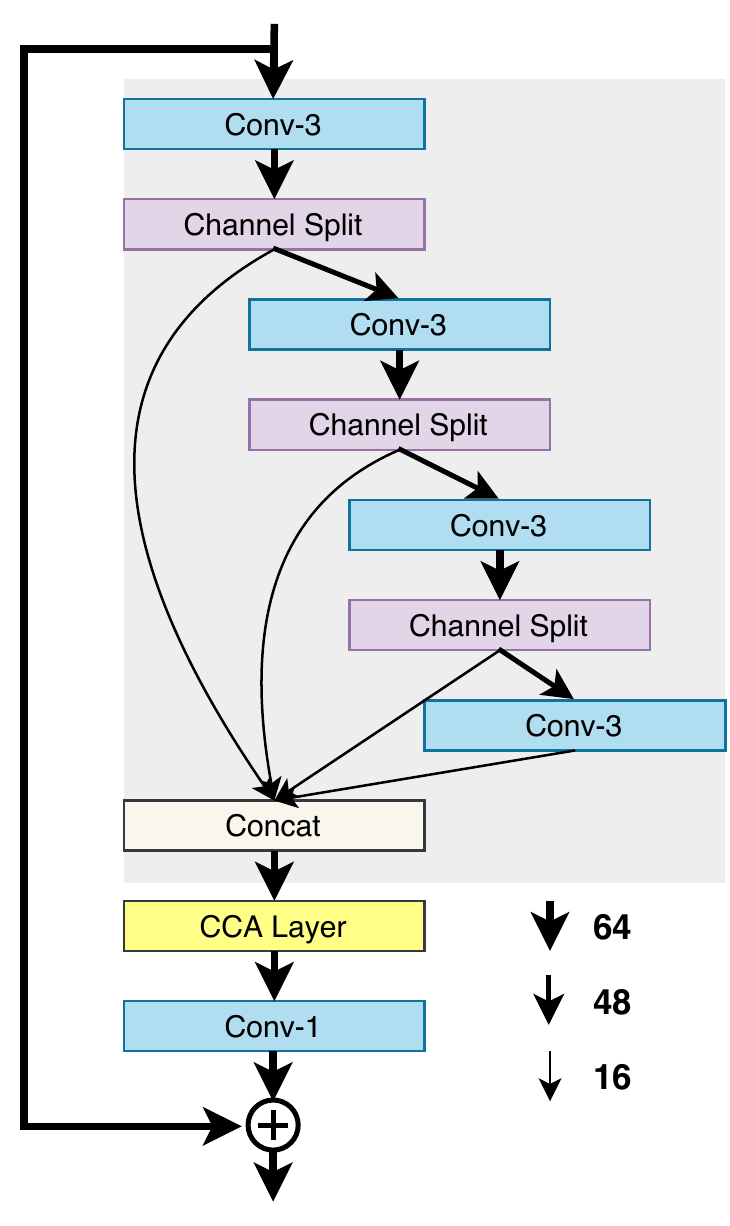}}
    \subfigure[]{\label{fig:imdb-r}\includegraphics[width=.272\linewidth]{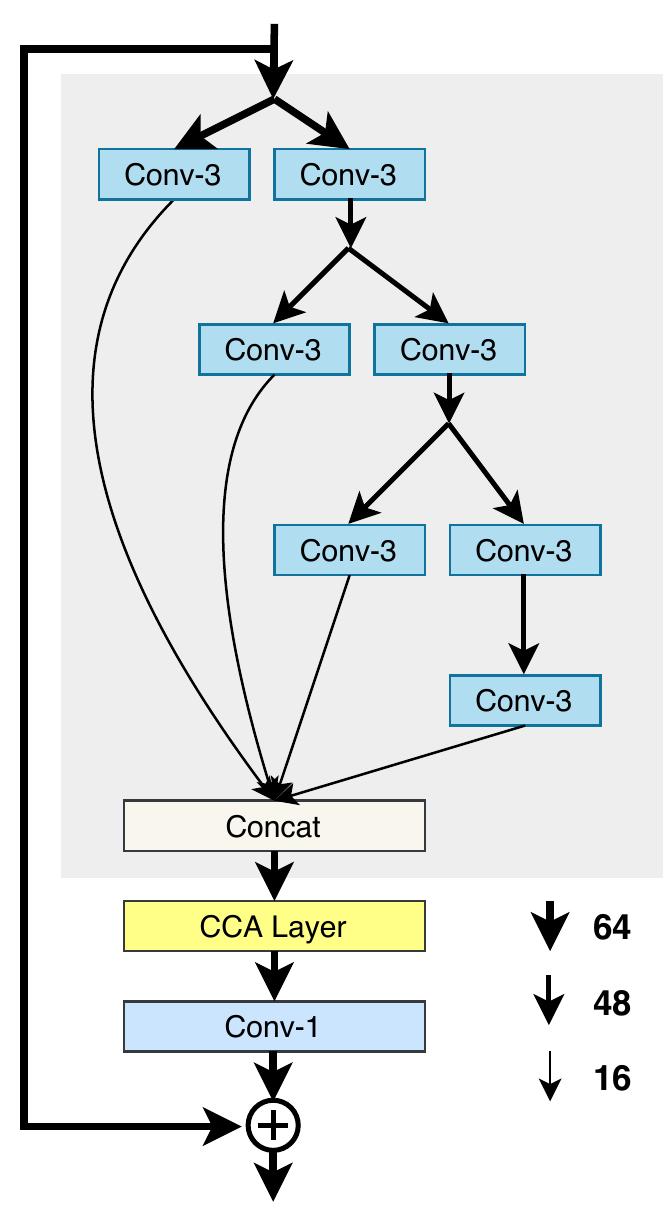}}
    \subfigure[]{\label{fig:rfdb}\includegraphics[width=.27\linewidth]{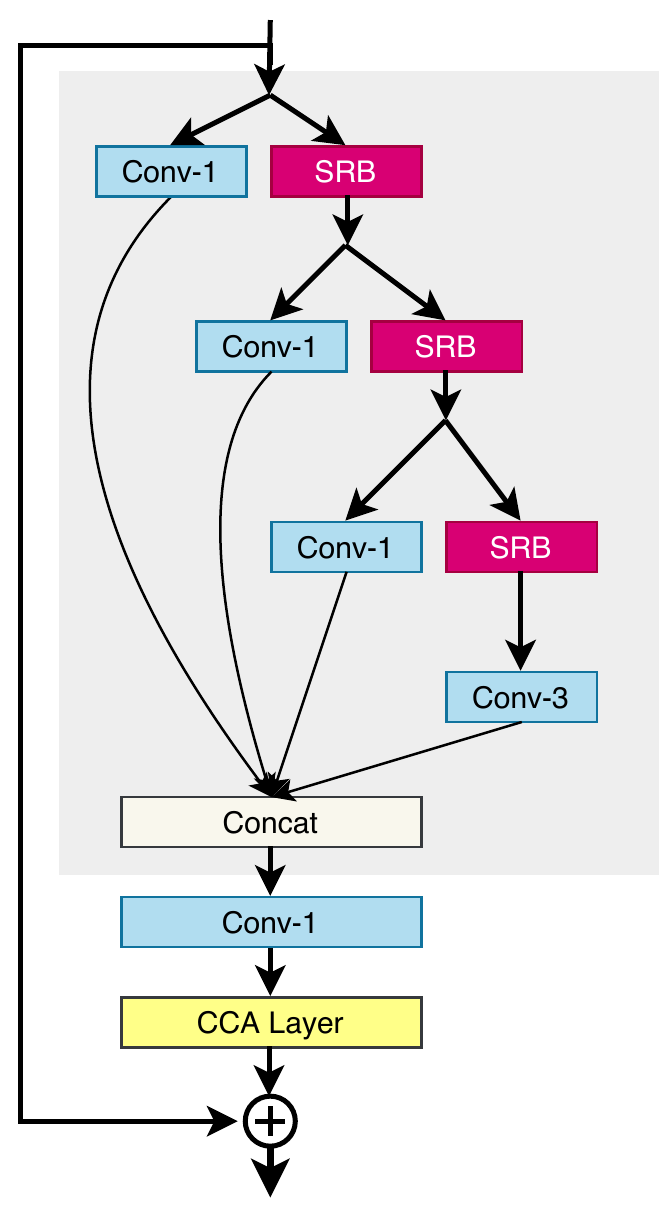}}
    \subfigure[]{\label{fig:srb}\includegraphics[width=.1\linewidth]{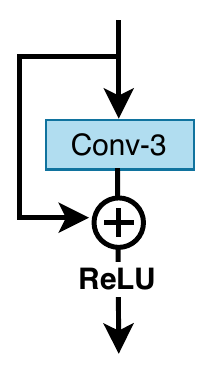}}
\caption{NJU\_MCG Team: (a) IMDB: the original information multi-distillation block. (b) IMDB-R: rethinking of the IMDB. (c) RFDB: residual feature distillation block. (d) SRB: shallow residual block.}
\label{fig:imdbs}
\end{figure}

The NJU\_MCG proposed \textbf{Residual Feature Distillation Network (RFDN)} for fast and lightweight image SR~\cite{liu2020residual}. The proposed RFDN is inspired by two recent works IMDN~\cite{hui2019lightweight} and RFANet~\cite{RFANet}.
As shown in Fig.~\ref{fig:imdb}, the main part of information distillation block (IMDB) is a progressive refinement module (PRM) marked with a gray background. 
Although PRM achieves prominent improvements, it is not efficient enough and introduces some inflexibility because of the channel splitting operation. The distilled features are generated by $3\times 3$ convolution filters that have many redundant parameters. Moreover, the feature refinement pipeline
(along the right branch of the PRM) is coupled together with channel splitting operation so that it is hard to use identity connections only for this pipeline. 
The IMDB-R in Fig.~\ref{fig:imdb-r} solves these problems by replacing the channel splitting operation with two concurrent $3\times 3$ convolutions. It is more flexible than the original IMDB. 
Based on this new architecture, RFDB in Fig.~\ref{fig:rfdb} uses three $1\times1$ convolutions for feature distillation. Furthermore, it uses the shallow residual block (SRB) in Fig.~\ref{fig:srb} as the feature extractor which can benefit most from the residual learning strategy. 
For shallow SR models, it is more efficient to use spatial attention than channel attention. So RFDB replaces the CCA layer with the ESA block in RFANet~\cite{RFANet}.

The proposed RFDN model contains 4 RFDBs, the overall framework follows the pipeline of IMDN, where global 
feature aggregation is used to augment the final features and the number of feature channels 
is set to 50. During the training of RFDN, HR patches of size $256\times 256$ are randomly cropped
from HR images, and the mini-batch size is set to 64. The RFDN model is trained by minimizing L1 loss
function with Adam optimizer. The initial learning rate is set to $5\times 10^{-4}$ and halved at every
200 epochs. After 1000 epochs, L2 loss is used for fine-tuning with learning rate of $1\times10^{-5}$. 
DIV2K and Flickr2K datasets are used for training the RFDN model.

\subsection*{AiRiA\_CG}
\begin{figure}[!h]
  \centering
  \includegraphics[width=.8\linewidth]{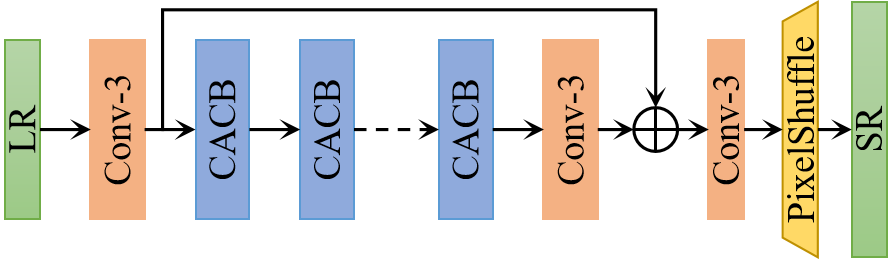}
  \caption{AiRiA\_CG Team: architecture of FIMDN.}
  \label{fig:IMDN}
\end{figure}

\begin{figure}[!h]
  \centering
  \includegraphics[width=0.8\linewidth]{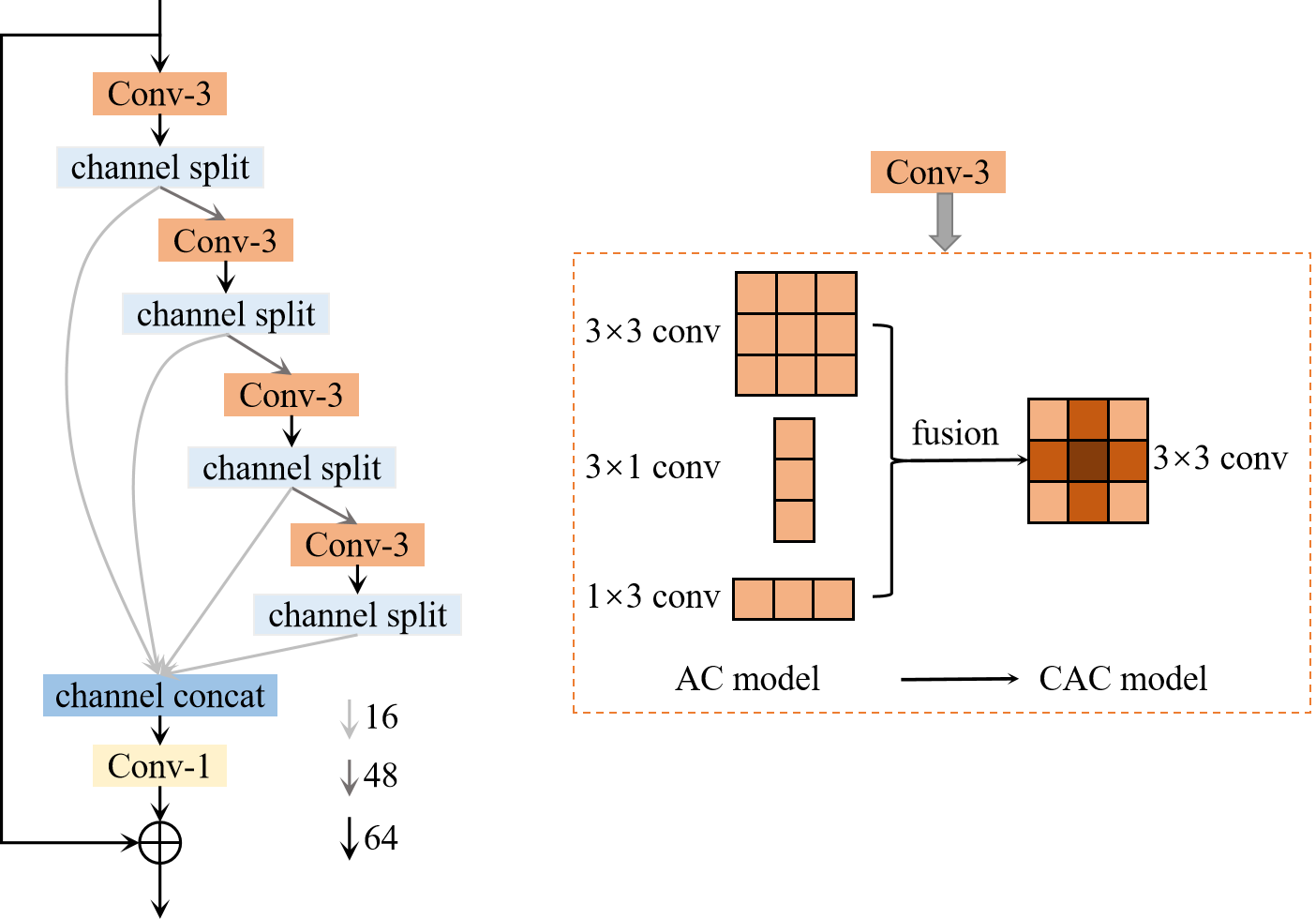}
  \caption{AiRiA\_CG Team: Left: Detailed architecture of FIMDN. ``Conv-1" denotes the $1\times1$ convlution layer. 64, 48, and 16 represent the number of output channels. Right: Details of the asymmetric $3\times3$ convolution. ``CAC" means converted Asymmetric Convolution (CAC).}
  \label{fig:CAC}
\end{figure}

The AiRiA\_CG team proposed \textbf{Faster Information Multi-Distillation Network via Asymmetric Convolution (FIMDN)}.
The proposed FIMDN shown in Fig.~\ref{fig:IMDN} is modified from IMDN~\cite{hui2019lightweight} with minor improvements such as less building blocks and converted asymmetric convolution. Different from IMDN, FIMDN only uses 6 CACBs as the building blocks to further accelerate the network. As illustrated in Fig.~\ref{fig:CAC}, FIMDN employs four `Conv-3' layers and retains a part of the information step-by-step. Then the hierarchical features are fused by using a $1\times1$ convolution layer. In particular, inspired by ACNet~\cite{Ding_2019_ICCV}, FIMDN utilizes the original AC model where $3\times3$ convolution layer is coupled with parallel $1\times3, 3\times1$ kernels. After the first training stage, the original AC model is converted into a single standard $3\times3$ convolution layer. Fig.~\ref{fig:CAC} illustrates the fusion process.
The training process contains two stages with four steps.
\begin{enumerate}
    \item At the first stage, the original AC model is equipped with three parallel asymmetric convolutions. 
    \begin{itemize}
        \item[I.] Pre-training on DIV2K+Flickr2K (DF2K). HR patches of size $256\times256$ are randomly cropped from HR images, and the mini-batch size is set to 64. The original FIMDN model is trained by minimizing L1 loss function with Adam optimizer. The initial learning rate is set to 2e-4 and halved at every 3600 epochs. The total number of epochs is 18000.
        \item[II.] Fine-tuning on DF2K.  HR patch size and the mini-batch size are set to $640\times640$ and 24, respectively. The FIMDN model is fine-tuned by minimizing L2 loss function. The initial learning rate is set to 1e-5 and halved at every 720 epochs. The total number of epochs is 3600.
        \item[III.] Fine-tuning on DIV2K. HR patch size and the mini-batch size are set to $640\times640$ and 4, respectively. The FIMDN model is finetuned by minimizing L2 loss function. The initial learning rate is set to 1e-6 and halved at every 400 epochs. The total number of epochs is 2000.
    \end{itemize}
    \item At the second stage, the final FIMDN model is obtained by converting three parallel convolutional kernels into a single $3\times3$ convolution layer. 
    \begin{itemize}
        \item[IV.] Fine-tuning on DIV2K. HR patch size is set to $640\times640$ and the mini-batch size is set to 24. The final CAC model is fine-tuned by minimizing L2 loss function. The initial learning rate is set to 1e-6 and halved at every 200 epochs. The total number of epochs is 1000.
    \end{itemize}
\end{enumerate}

\subsection*{UESTC-MediaLab}
\begin{figure}[!ht]
   \centering
   \includegraphics[width=0.6\linewidth]{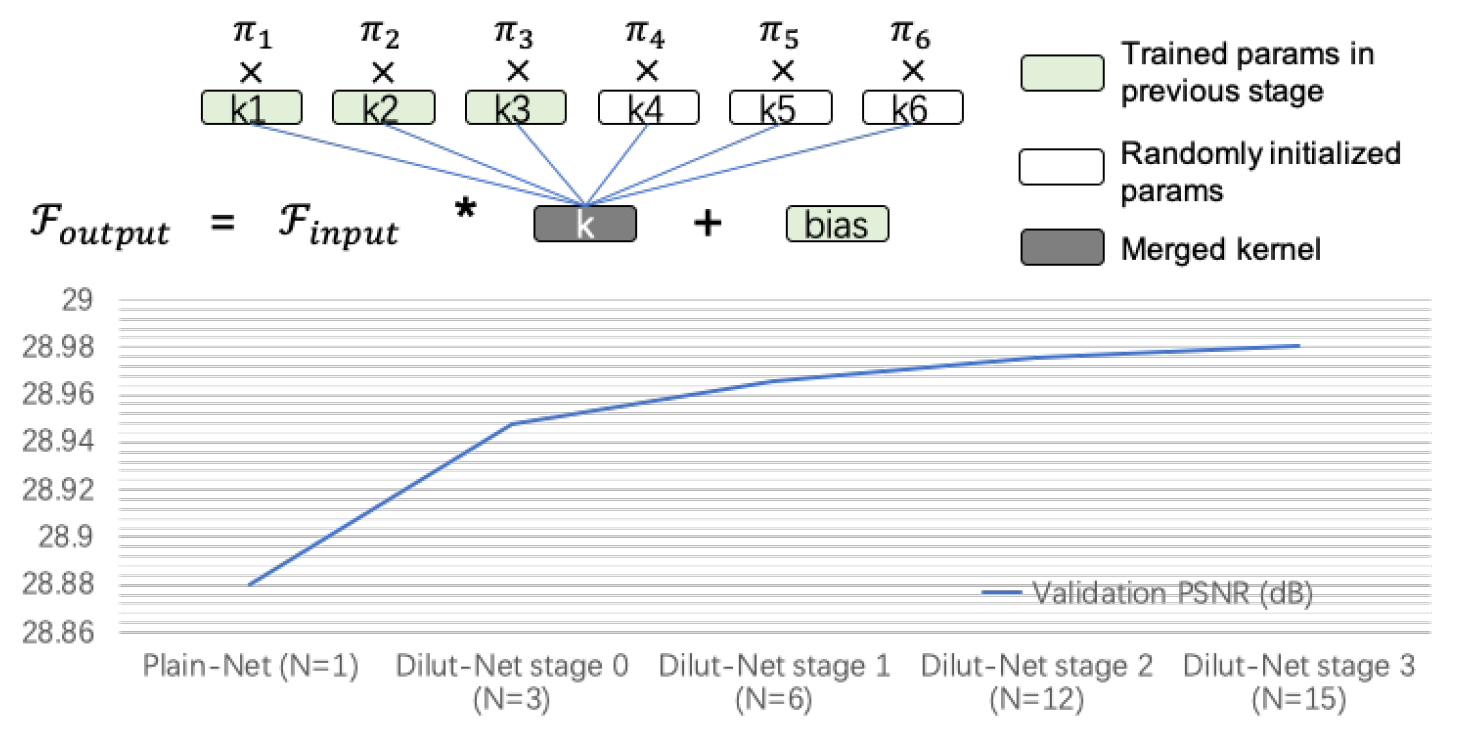}
   \caption{UESTC-MediaLab Team: kernel dilution and performance of different training stages.}
   \label{fig:UESTC-MediaLab}
\end{figure}

The UESTC-MediaLab team proposed a novel training strategy which is able to boost the performance of CNNs without extra parameters. The traditional convolution layer can be formulated as $\mathbf{F}_{output} = \mathbf{F}_{input} * \tilde{\mathbf{k}} + \mathbf{b}$. The team decomposed the kernel $\tilde{\mathbf{k}}$ into $N$ kernel bases, \ie $\tilde{\mathbf{k}} = \sum_{i}{\pi_i \times \mathbf{k}_i}$, where $\{\pi_i | i = 1,2,\dots,N\}$ are trainable merging weights~\cite{li2019learning}. 
The proposed training procedure has multiple stages.
At the first stage (denoted as $0$-th stage), the model is trained from scratch with the number of kernel bases $N_0 = 3$ in each layer. All kernel bases are initialized randomly with Xavier-Uniform \cite{Glorot2010Understanding}, merging weights with $\frac{1}{N_0}$ and bias with zeros. 
At the $t$-th ($t \geq 1$) stage, $\delta_t$ kernel bases in each layer which are added and randomly initialized, and the merge weights are initialized with $\frac{1}{N_t}$ where $N_t = N_{t-1} + \delta_t$. 
Each training stage terminates when the validation loss converges. After the training phase, only merged kernels $\tilde{\mathbf{k}}$ and the bias $\mathbf{b}$ are saved. The kernel bases $\mathbf{k}_i$ and merging weights $\pi_i$ are not necessary in the inference phase. As shown in Fig.~\ref{fig:UESTC-MediaLab}, the performance increases gradually with the number of kernel bases in Dilut-Net (the network trained with the proposed strategy). The Plain-Net denotes the network using traditional convolutional layers without kernel bases.

As for the model structure, the UESTC-MediaLab team modified the structure of IMDN~\cite{hui2019lightweight} from the following aspects.
Firstly, the computing unit `conv3$\times$3 $\rightarrow$ LeakyRelu' is replaced by `gconv3$\times$3 $\rightarrow$ PRelu $\rightarrow$ conv1$\times$1', where gconv denotes the group convolution which doubles the number of feature maps. Secondly, adaptive skip-connection is adopted in the model, which parameterizes the connection relationship between the outputs of blocks. Thirdly, the depth and width of the network is modified to achieve a better balance between efficiency and effectiveness.

\subsection*{XPixel}
\begin{figure}[!ht]
    \centering
    \includegraphics[width=.8\linewidth]{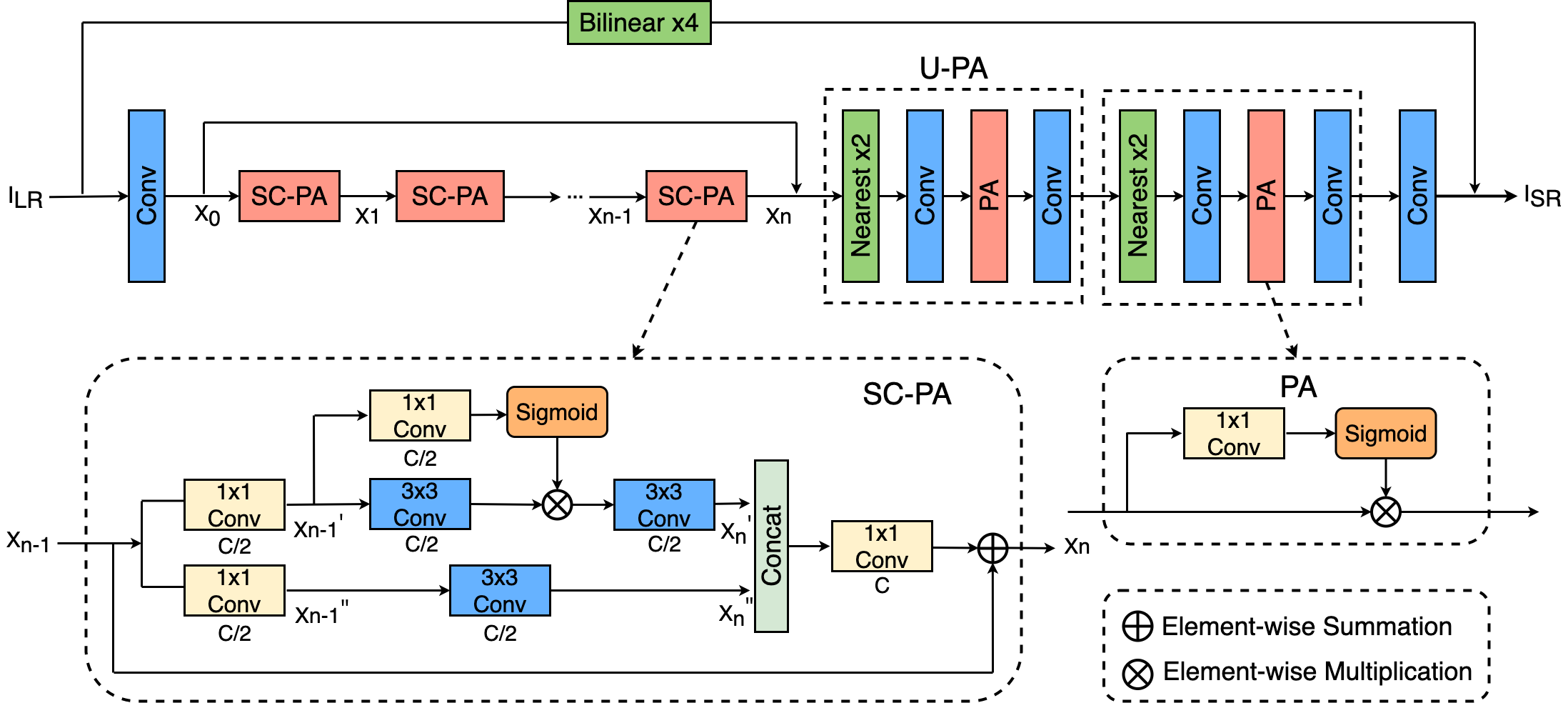}
    \caption{XPixel Team: the network architecture.}
    \label{fig:XPixel}
 \end{figure}

The XPixel team proposed \textbf{a simple yet efficient network called Pixel Attention Network (PAN)}~\cite{zhao2020efficient}, which achieves the same performance as MSRResNet with only 272,419 parameters. The main contribution is ``Pixel Attention'' scheme.
The framework consists of three stages depicted in Fig.~\ref{fig:XPixel}. First, one  convolution layer extracts the features, then 16 SC-PA blocks are utilized for non-linear mapping, at last there are two U-PA blocks for final reconstruction. 
Pixel attention is adopted in the SC-PA and U-PA blocks. 
Next, the details of these new modules will be described.

\textbf{Pixel Attention.} 
First, let us revisit channel attention~\cite{CSE} and spatial attention~\cite{SSE}. 
Channel attention aims to obtain 1D ($C \times 1 \times1$) attention feature vector, while spatial attention obtains a 2D ($1 \times H \times W$) attention map. Note that $C$ is the number of channels, $H$ and $W$ are the height and width of the features, respectively. Different from them, proposed pixel attention is able to generate a 3D  ($C \times H \times W$) matrix as the attention features. As shown in Fig.~\ref{fig:XPixel}, pixel attention only uses a $1\times1$ convolution layer and a sigmoid function to obtain the attention maps which will then be multiplied with the input features.

\textbf{SC-PA Block.} The main block in the network is called Self-Calibrated convolution with Pixel Attention (SC-PA). 
Specifically, proposed pixel attention scheme is added to the Self-Calibrated convolution module. 
Basically, SC-PA is comprised of two branches. Each branch contains a $1\times1$ convolution layer at the beginning, which will reduce half of the channel number. 
The upper branch also contains two $3\times3$ convolution layers, where the first one is equipped with a PA module.
This branch transforms $X_1$ to $Y_1$.
In the second branch, only a single $3\times3$ convolution layer is used to generate $Y_2$ since the original information should be maintained. Finally, $Y_1$ and $Y_2$ are concatenated into $Y_3$, which will then be passed to a $1\times1$ convolution layer. In order to accelerate training, shortcut is used to produce the final output features $Y$.

\textbf{U-PA Block.} Except for the main blocks, pixel attention is also adopted in the final reconstruction stage.
Specifically, proposed U-PA block is added after each upsampling layer. 
Note that in previous SR networks, a reconstruction stage is basically comprised of upsampling and convolution layers.
Besides, plenty of SR networks (e.g. MSRResNet) use PixelShuffle layer to upsample the features, which is computational expensive. 
To reduce the computation cost and parameters in the reconstruction stage, Nearest Neighbor interpolation layer is used to replace the PixelShuffle layer and the following convolution layer will reduce the number of channels by half. 
Moreover, previous works have shown that attention mechanism can effectively improve the performance in SR tasks but few researchers investigate it in the upsampling stage.
In this work, the U-PA block is introduced after each upsampling layers.
As shown in Fig.~\ref{fig:XPixel}, the U-PA block consists of a PA layer between two convolution layers. Experiments have been conducted to demonstrate its effectiveness.

\textbf{Implementation details.} During training, DIV2K and Flickr2K are uesd as training datasets. The HR patch size is set to $256\times256$. The batch size is set to 32. L1 loss function is adopted with Adam optimizer to train the model. During validation, the model achieves an average PSNR of 29.00 dB on DIV2K validation dataset. The inference time is about 0.0758s per image with a single GTX 1080Ti GPU.

\subsection*{HaiYun}
\begin{figure}[!ht]
    \centering
    \includegraphics[width=.8\linewidth]{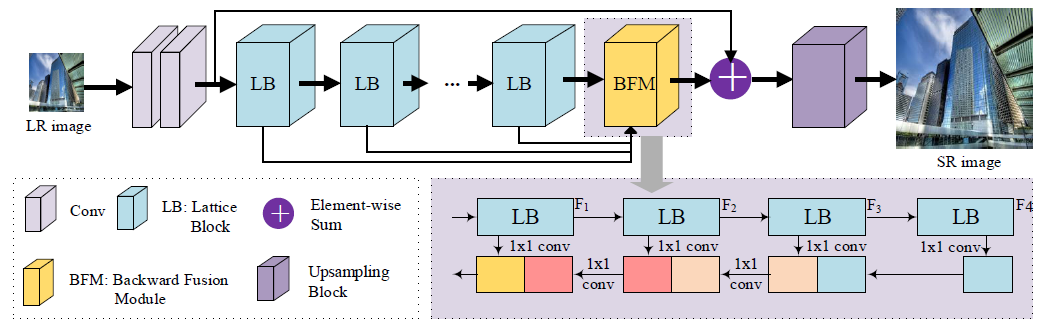}
    \caption{HaiYun Team: the network architecture.}
    \label{fig:HaiYun}
 \end{figure}

The HaiYun team proposed \textbf{a lightweight SR network} (see Fig.~\ref{fig:HaiYun}). Due to the frequent use of residual block (RB) in SR models, they pursue an economical structure to adaptively combine RBs. Inspired by lattice filter bank, a lattice block (LB) is designed where two butterfly structures are applied to combine two RBs. LB has the potential of various linear combinations of two RBs. Each case of LB depends on the combination coefficients which are determined by the attention mechanism. LB favors the lightweight SR model with the reduction of about half amount of the parameters while keeping the similar SR performance. Moreover, a lightweight SR model, \ie LatticeNet, is proposed, which uses a series of LBs and the backward feature fusion.

\subsection*{IPCV\_IITM}
\begin{figure}[!ht]
    \centering
    \includegraphics[width=.8\linewidth]{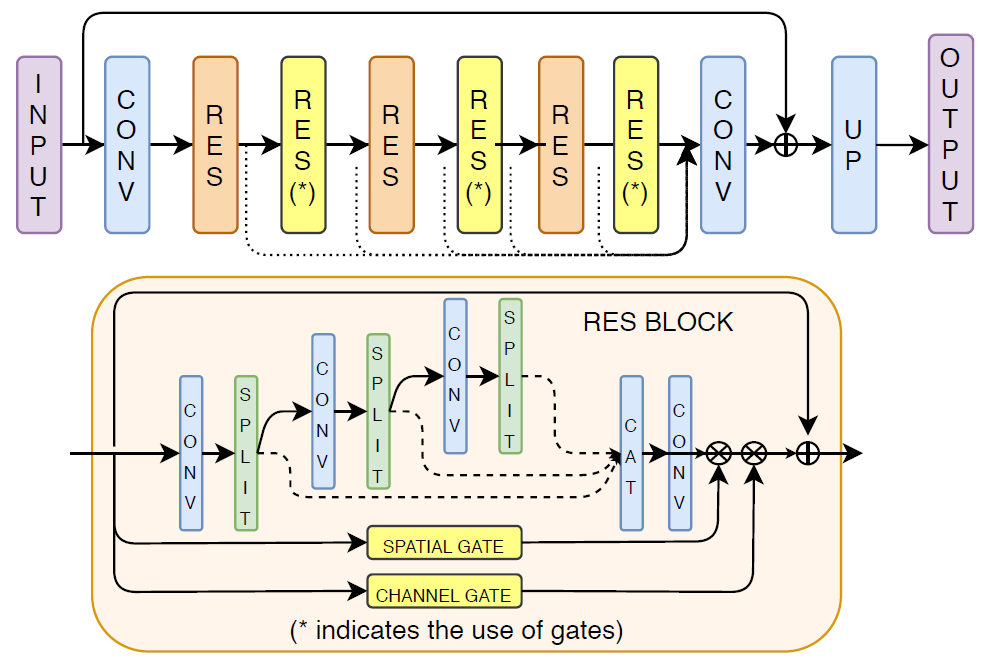}
    \caption{IPCV\_IITM Team: the network architecture.}
    \label{fig:IPCV}
 \end{figure}

The IPCV\_IITM team proposed to \textbf{stack multiple residual blocks for gradual feature refinement}. One $3\times3$ convolution with 64 output channels is used to extract features from the LR image. Multiple residual blocks are stacked together and at the end all intermediate feature maps are fused by a $1\times1$ convolution layer. 
As shown in Fig.~\ref{fig:IPCV}, residual blocks perform channel split operation on the preceding features, producing two feature splits. One is preserved and the other portion is fed into the next calculation unit. The retained part is used as the refined features. 
Each convolutional operation is followed by a Leaky ReLU activation function
except for the last $1\times1$ convolution. Two sub-networks are used for predicting spatial and channel attention map to perform feature-wise gating operation on the features. For the purpose of faster inference, the gating
function is used in alternating residual blocks as indicated in the figure. Given a feature map of size $C\times H\times W$, the spatial and channel attention module produces $H\times W$ and $C$ dimensional soft gates, which are element-wise multiplied with the feature. The same sub-networks are used for all the residual blocks to reduce the number of parameters. At the end, the gated feature is added to the input. The up-sampler at the end of the network includes one $3\times3$ convolution and a sub-pixel convolution.

\subsection*{404NotFound}
\begin{figure}[!ht]
    \centering
    \includegraphics[width=.8\linewidth]{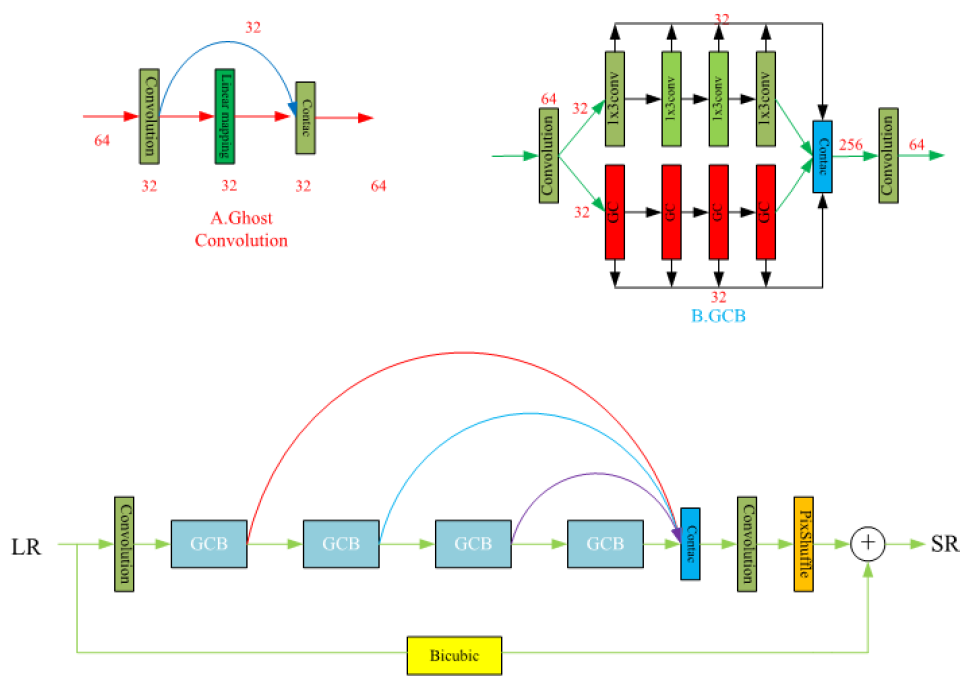}
    \caption{404NotFound Team: the network architecture.}
    \label{fig:404NotFound}
 \end{figure}

The 404NotFound team proposed \textbf{GCSR} which includes
three main parts (see Fig.~\ref{fig:404NotFound}). Firstly, the input channels are divided into 2 parts. Ghost convolution and $1\times3$ convolution are adopted to replace normal convolution. Secondly, a special loss is proposed which consists of L1 loss for low frequency information reconstruction and gradient loss for high-frequency information reconstruction. Thirdly, bicubic upscaling for the LR image is used.

\subsection*{MDISL-lab}
The MDISL-lab team proposed \textbf{PFSNet}.
The main idea is to keep the width of features in an SR network by making full use of a shared feature in every block for efficient SR. In details, the MDISL-lab team found that keeping the width of features is necessary for the high performance of SR networks. And there are many redundancy in feature maps of SR networks. Thus, a series of cheap operations with cheap cost are applied to a shared feature to generate many simple features. These simple features are then concatenated with some normal features at the channel dimension and fed to convolutional layers. In this way, the width of output feature of most convolutional layers could be reduced, thus reducing the computational cost. Meanwhile, the width of input features could still be maintained by the shared feature followed by cheap operations.

\begin{figure}[!ht]
   \centering
   \includegraphics[width=0.7\linewidth]{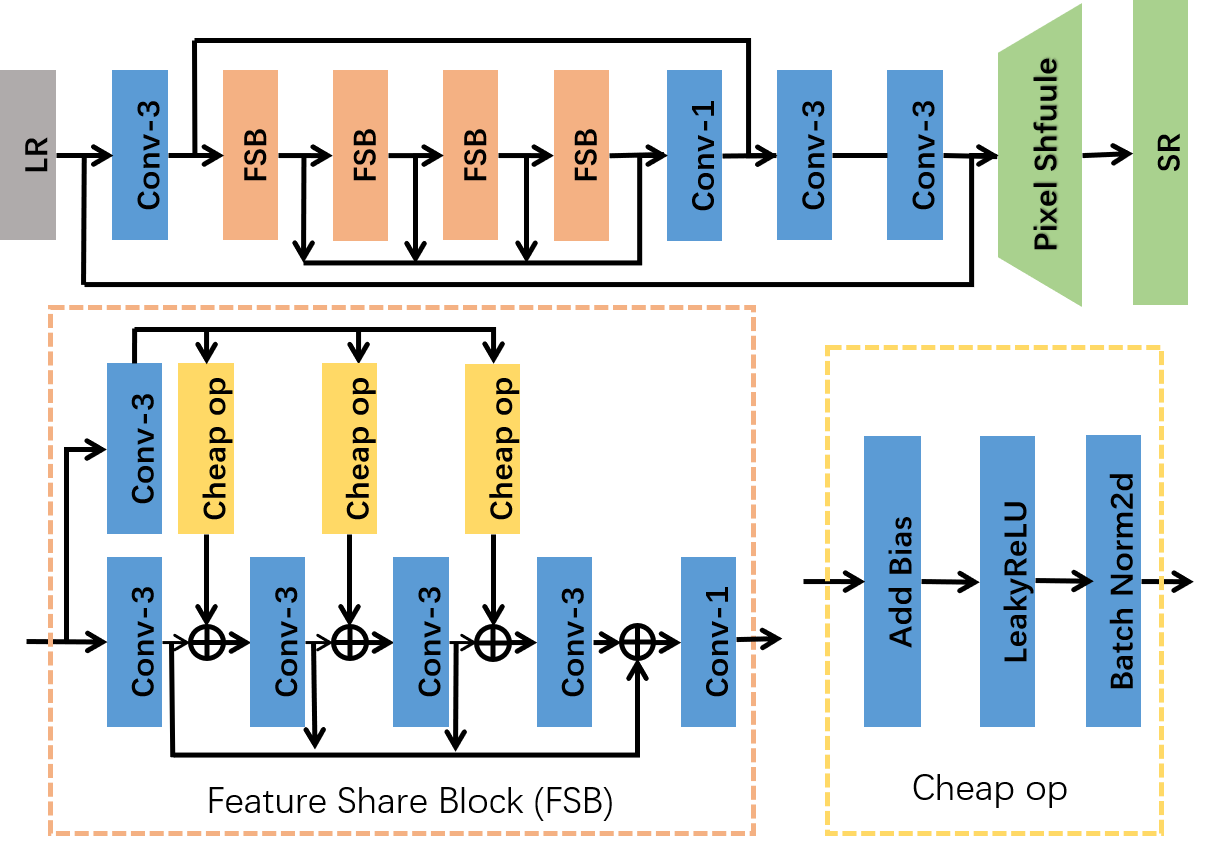}
   \caption{MDISL-lab Team: the network architecture.}
   \label{fig:MDISL-lab}
\end{figure}

The architecture of PFSNet is shown in Fig.~\ref{fig:MDISL-lab}. First, the Feautre Share Block (FSB) is constructed by several $3\times 3$ convolutional layers, a shared $3\times3$ convolutional layer, several cheap operations. And a $1\times 1$ convolutional layer is used to reduce the number of feature channels. The $\mathbf{\oplus}$ symbol in the figure represents concatenation at channel dimension. All $3\times3$ convolutional layers output features with 32 channels while the input features of all convolutional layers has 64 channels.

\subsection*{MLVC}
\begin{figure}[!ht]
  \centering
  \includegraphics[width=.8\linewidth]{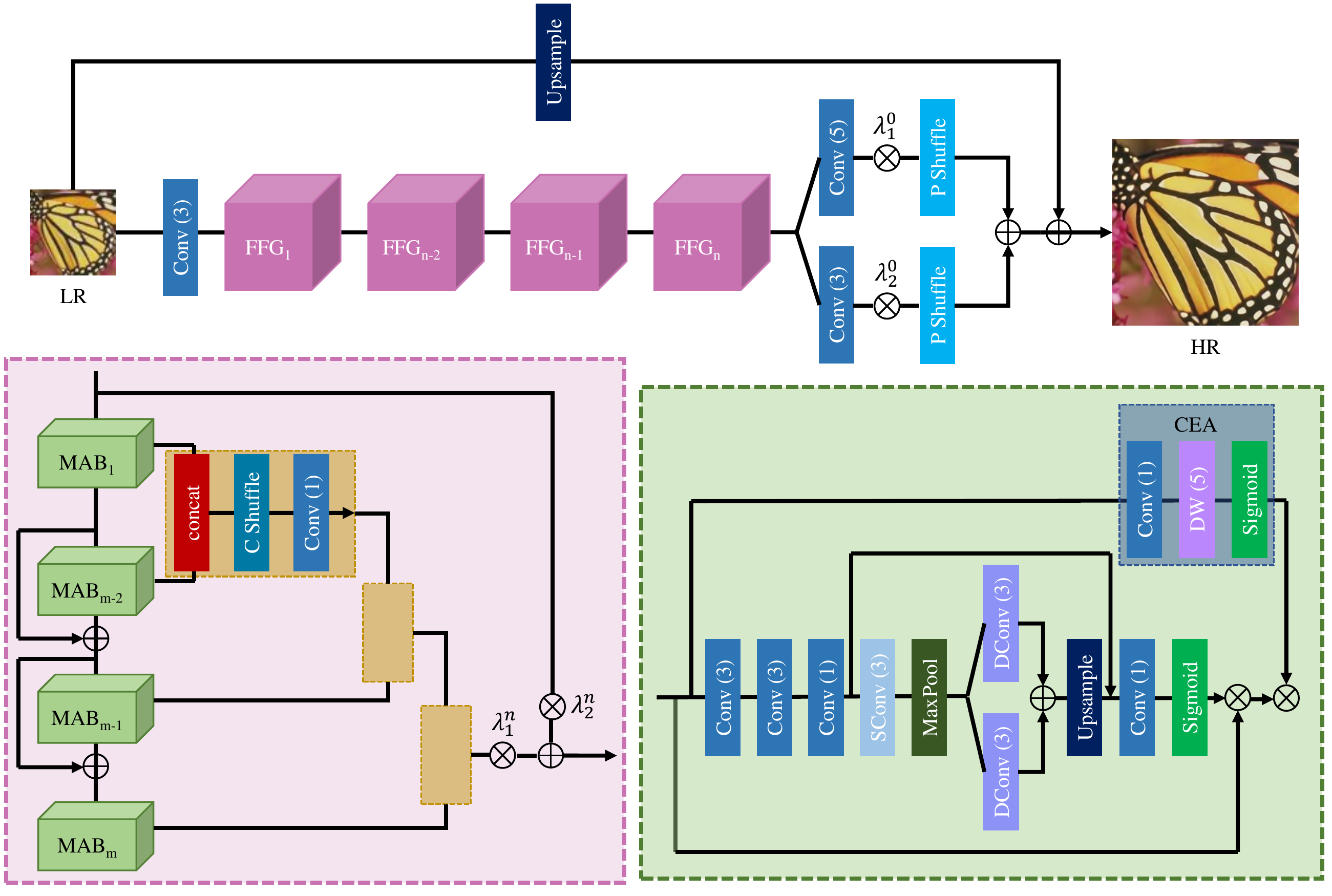}
  \caption{MLVC Team: the network architecture. The top image represents the whole structure of proposed model, bottom left represents structure of feature fusion group (FFG), and bottom right shows structure of multi-attention block (MAB).}
  \label{fig:main_fig}
\end{figure}

The MLVC proposed \textbf{Multi-Attentive Feature Fusion Super-Resolution Network (MAFFSRN)}~\cite{muqeet2020ultra}. The main architecture shown in Fig.~\ref{fig:main_fig} is based on RDN~\cite{zhang2018residual} that consists of local and global blocks. The feature fusion group (FFG) and multi-attention block (MAB) are used as global and local blocks, respectively. The MAB is inspired by enhanced spatial attention (ESA)~\cite{RFANet}. MAB introduces another cost-efficient attention mechanism (CEA)~\cite{cai2020learning} to refine the input features. The CEA basically consists of point-wise and depth-wise convolutions. It is incorporated into the MAB block to improve the performance of the network with negligible additional computational cost. Furthermore, it is observed that the original ESA block is computationally expensive due to convolutional group. Thus, the convolutional group is replaced by dilated convolutions with different dilated factors. Lastly, element-wise addition is performed between the output of dilated convolutions to avoid the grid effects~\cite{yu2017dilated}. FFG is composed of a stack of MAB which are combined using binarized feature fusion (BFF)~\cite{muqeet2019hybrid}.

\subsection*{XMUlab}
\begin{figure}[!ht]
    \centering
    \includegraphics[width=.8\linewidth]{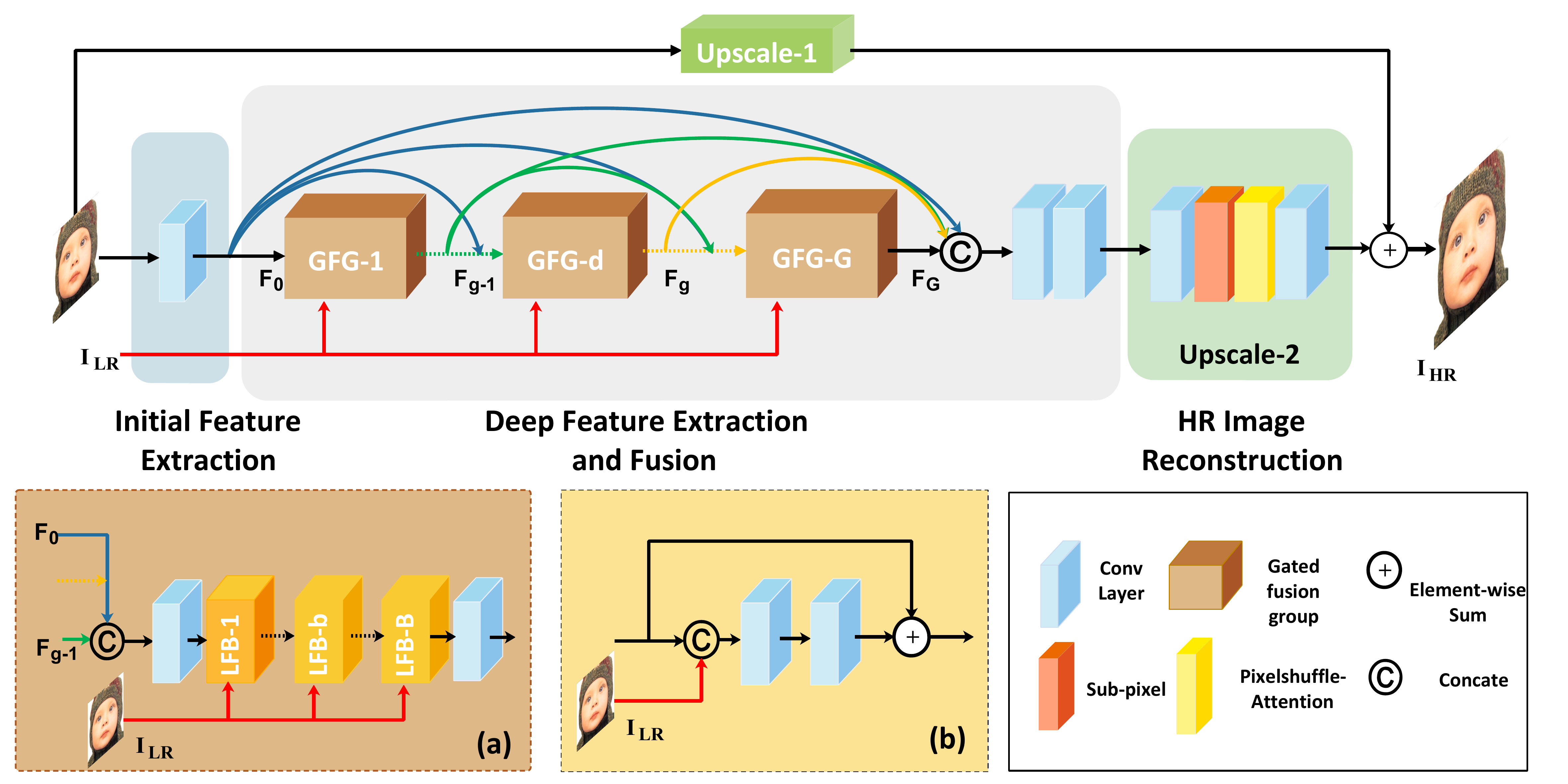}
    \caption{XMUlab Team: the network architecture.}
    \label{fig:XMUlab}
\end{figure}

The XMUlab team proposed \textbf{PixelShuffle Attention Network}. As shown in Fig.~\ref{fig:XMUlab}, the network contains several gated fusion groups. Each group has 4 residual blocks. Unlike the common use of attention module, PixelShuffle attention Network applies spatial attention and channel attention to the upscaled feature. Gated fusion group gather the output features from previous groups. The channel number in the network is reduced from 64 to 32 compared with other methods, which enables the network goes deeper. The LR image is also concatenated with feature map to extract more information.

\subsection*{MCML-Yonsei}
\begin{figure}[!ht]
    \centering
    \includegraphics[width=.8\linewidth]{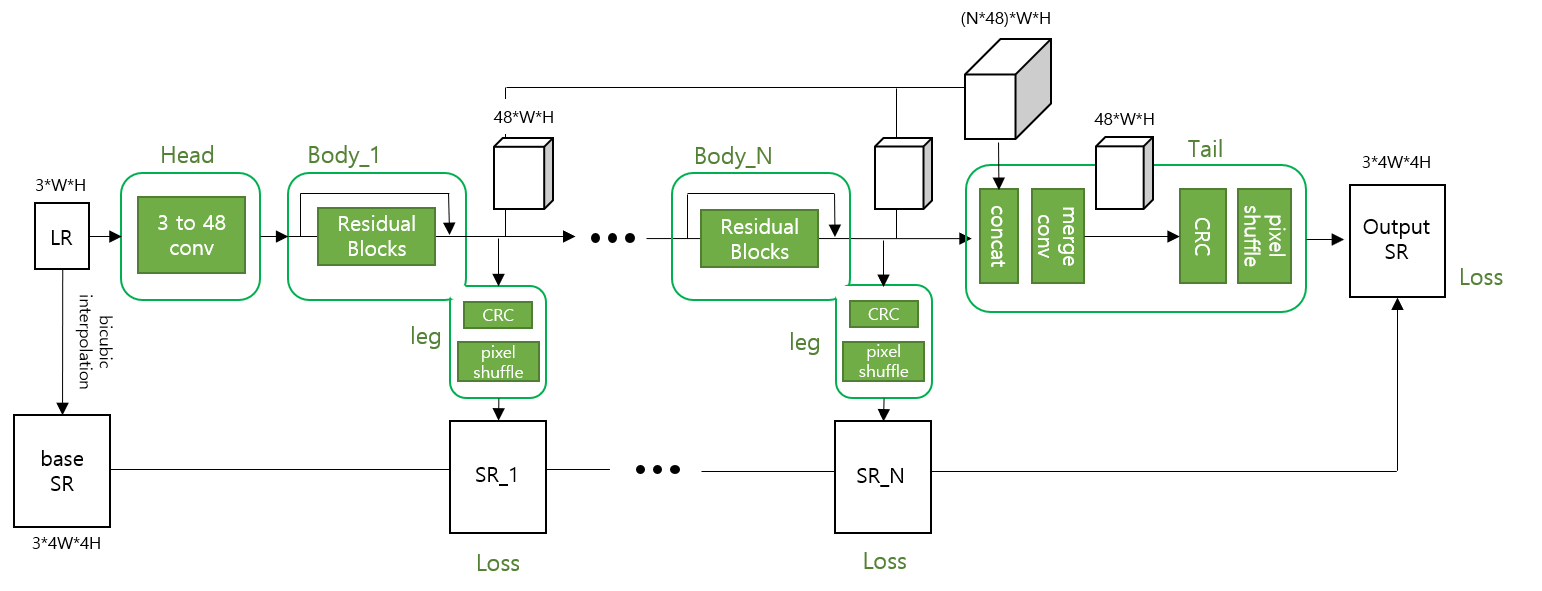}
    \caption{MCML-Yonsei Team: the network architecture.}
    \label{fig:LarvaNet}
\end{figure}

The MCML-Yonsei proposed \textbf{LarvaNet: Hierarchical Super-Resolution via Multi-exit Architecture}~\cite{jeon2020larvanet} (see Fig.~\ref{fig:LarvaNet}).  The main idea of the proposed LarvaNet is to divide a network into some body modules and to make each body module to generate an output SR image. This is inspired by
the fact that MSRResNet is effective for generating residual contents
which is added to the image interpolated from the LR image. In their
experiment, using interpolation increases PSNR by about 0.2 dB
compared to the MSRResNet model without interpolation. Like MSRResNet
generates residual contents to be added with a interpolated image,
each body module of LarvaNet generates residual contents to be added
with the previous module’s output image. With this idea, the model can
learn features important for SR at early stages, and can generate a
more accurate SR image because of the accumulation of residual
information.

In order to search for an optimal base architecture, they investigate
extensive variations of the baseline MSRResNet model in terms of the
numbers of channels and blocks, upsampling methods, interpolation
methods, weight initialization methods, and activation functions.
Following the results of the investigating experiments, they used a
MSRResNet with best settings as base architecture of the multi-exit
architecture, LarvaNet. The overall architecture of LarvaNet is
illustrated in Fig. 12. Each body module consisting of residual blocks
generates features and the previous features are accumulated by skip
connection. A sub-module is appended to the body module. It generates
an SR output image using features from the body module and
interpolated SR image. The tail module takes all the features
generated by the body modules and concatenates the features, and
generates a final output image. The average of all the losses from the
early and final output images is considered as the loss function of
training the model.

\subsection*{LMSR}
\begin{figure}[!ht]
    \begin{center}
        \includegraphics[width=.8\linewidth]{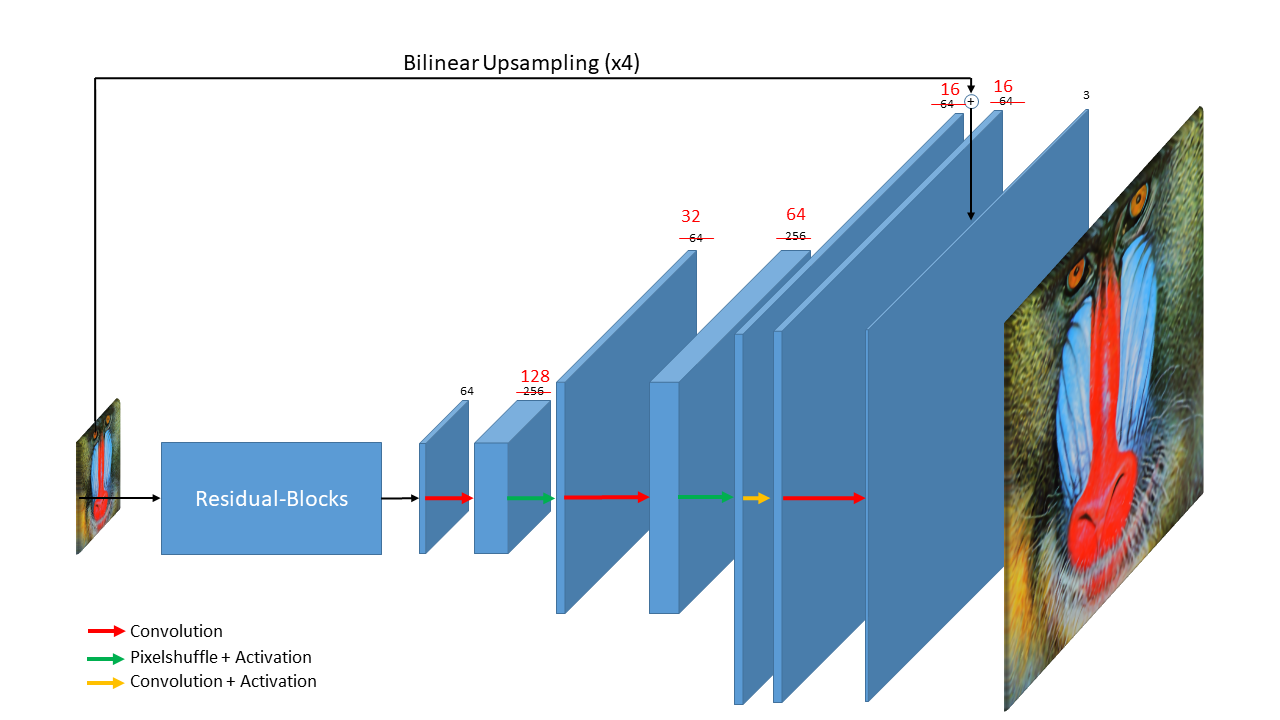}
        \caption[Lightweight Upsampling]{LMSR Team: the comparison between the lightweight upsampling method (red) and the one of the baseline model. The numbers represent the number of channels in each layer.}
        \label{fig:LMSR}
    \end{center}
\end{figure}

LMSR adapts the MSRResNet.
First, the activation function is replaced by the MTLU~\cite{gu2018multi}, which adds complexity to the model, while only making runtime measurements slightly worse.
On top of that, the complexity of the upsampling part of the MSRResNet is reduced, by only doubling the numbers of channels before doing a pixel-shuffle compared to quadrupling them as in the baseline model (Fig.~\ref{fig:LMSR}). In this way, a large number of operations can be reduced. Those convolutions contain the heaviest computations as the spatial resolution has a quadratic influence on the number of operations.
Finally, the number of residual blocks is reduced to 13.

\subsection*{ZJUESR2020}
\begin{figure}[!ht]
\centering
\includegraphics[width=0.7\textwidth]{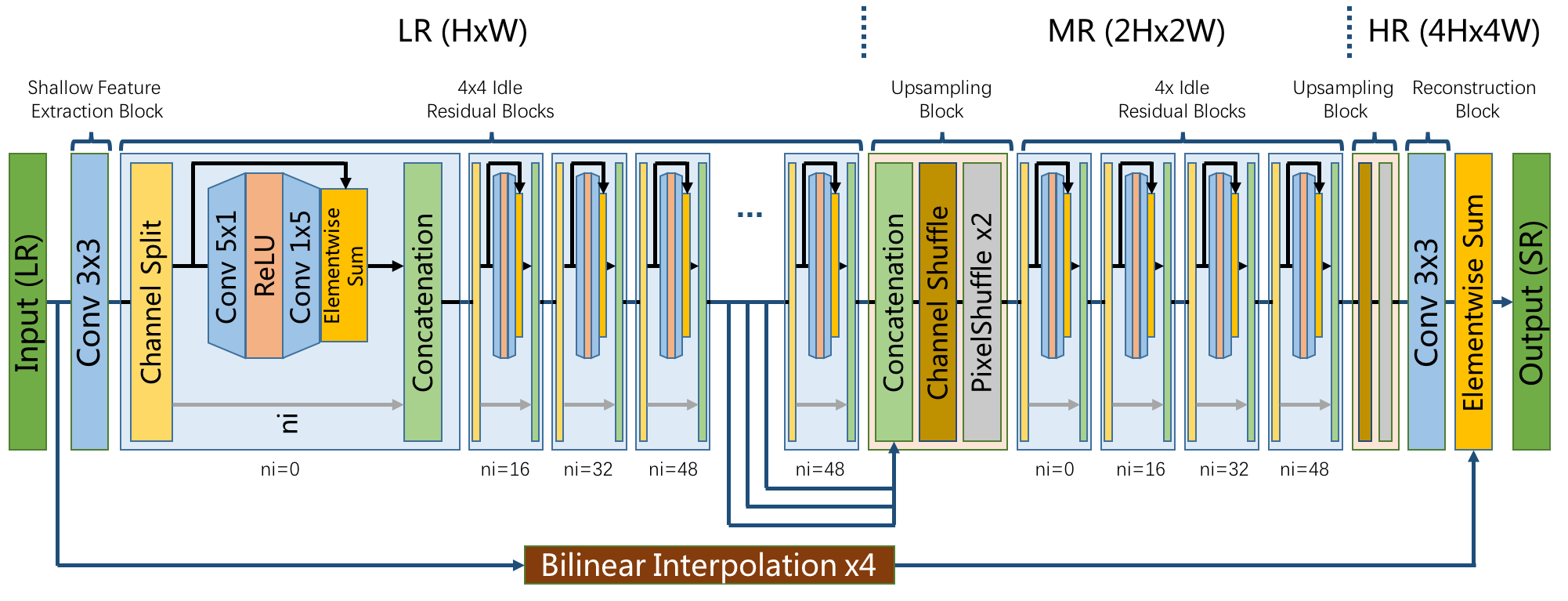}
\caption{ZJUESR2020 Team: the network architecture.}
\label{fig:idleSR}
\end{figure}

The ZJUESR2020 team proposed \textbf{IdleSR}~\cite{xiong2020aimesr} (see Fig.~\ref{fig:idleSR}). The basic architecture of IdleSR is same to that of NoUCSR~\cite{Xiong19}. To achieve a better trade-off among performance, parameters, and inference runtime, IdleSR adopts the following three strategies. 1) Multi-scale IdleBlocks are proposed to extract hierarchical features at the granularity of residual block. 2) Asymmetric kernels are applied to the residual blocks, which reduces nearly half of parameters and operations while keeping the receptive field the same. 3) Gradient scaling, larger
LR patch size (64), and longer training time (2M iterations) are used to compensate for the dropped performance during training phase.

\begin{figure}[!ht]
\centering
\subfigure[IMDB~\cite{hui2019lightweight}] {
	\includegraphics[width=0.25\textwidth]{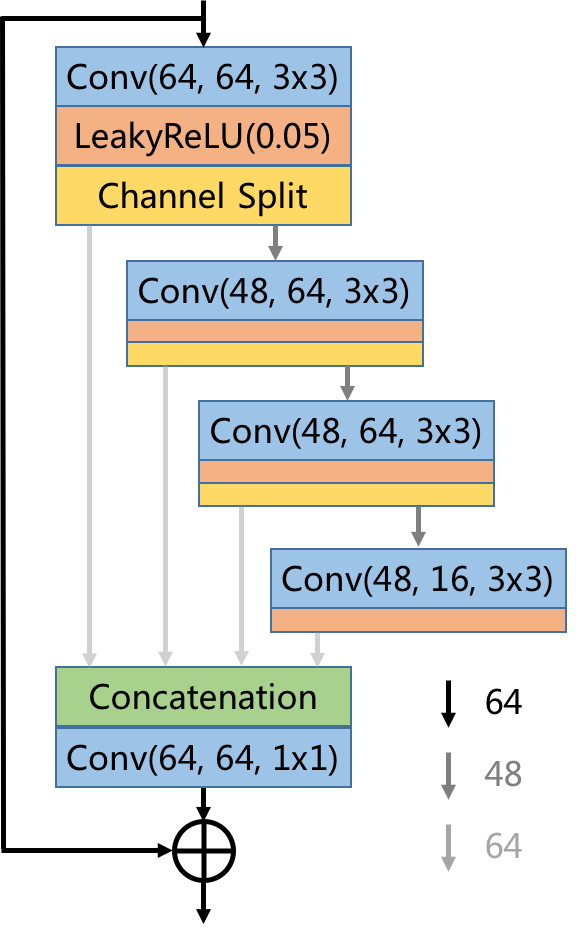}
	\label{fig:imdblock}
}
\subfigure[IdleBlock~\cite{Xu19}] {
	\includegraphics[width=0.25\textwidth]{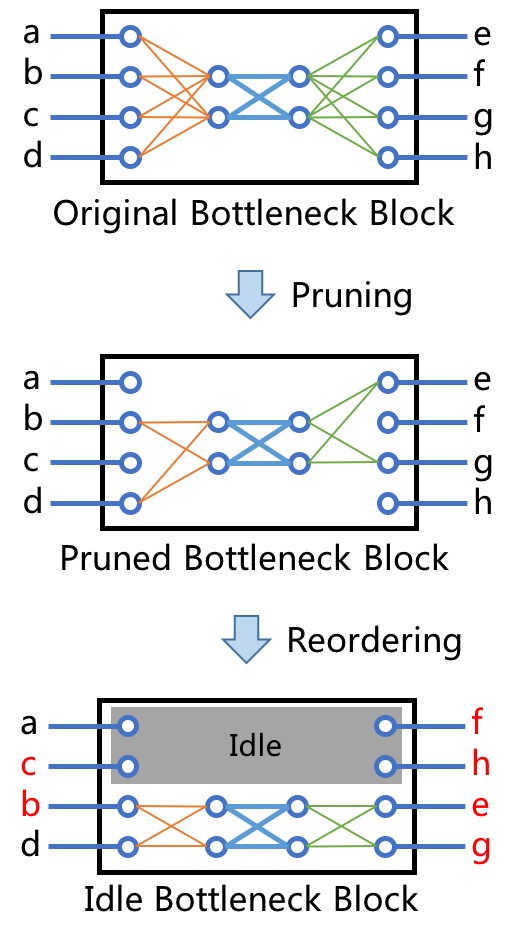}
	\label{fig:idleblock}
}
\subfigure[Multi-Scale IdleBlocks] {
	\includegraphics[width=0.25\textwidth]{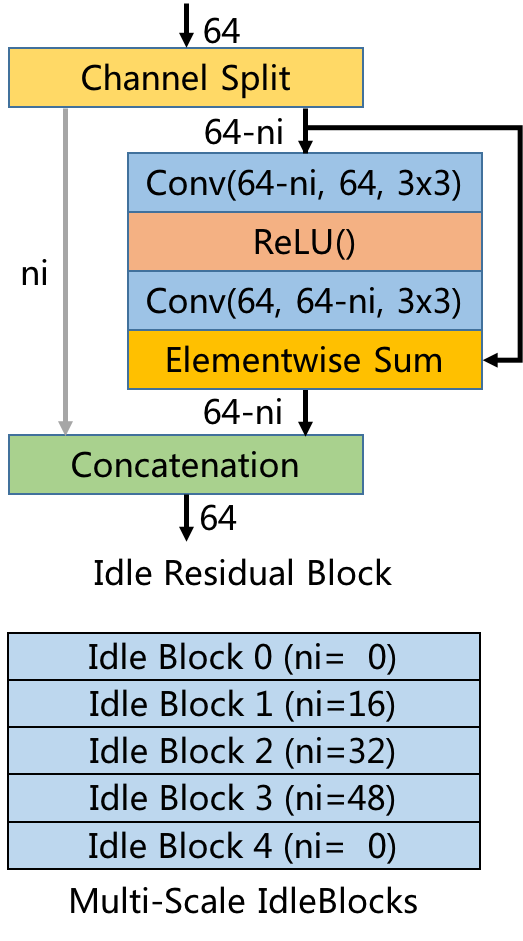}
	\label{fig:msidleblock}
}
\caption{ZJUESR2020 Team: the architecture of IMDB, IdleBlock and proposed Multi-Scale IdleBlocks.}
\label{fig:blocks}
\end{figure}

Multi-scale IdleBlocks combine the advantage of Information Multi-Distillation Block (IMDB)~\cite{hui2019lightweight} and IdleBlocks~\cite{Xu19}. Fig.~\ref{fig:blocks} compares the architecture of the three blocks.
Compared to IMDB, multi-scale IdleBlocks can avoid the usage of bottleneck convolution layer and reduce the amount of channel split operations. 

\subsection*{SC-CVLAB}
\begin{figure}[!ht]
\centering
\subfigure[]{\label{fig:HB_arch}\includegraphics[width=.32\linewidth]{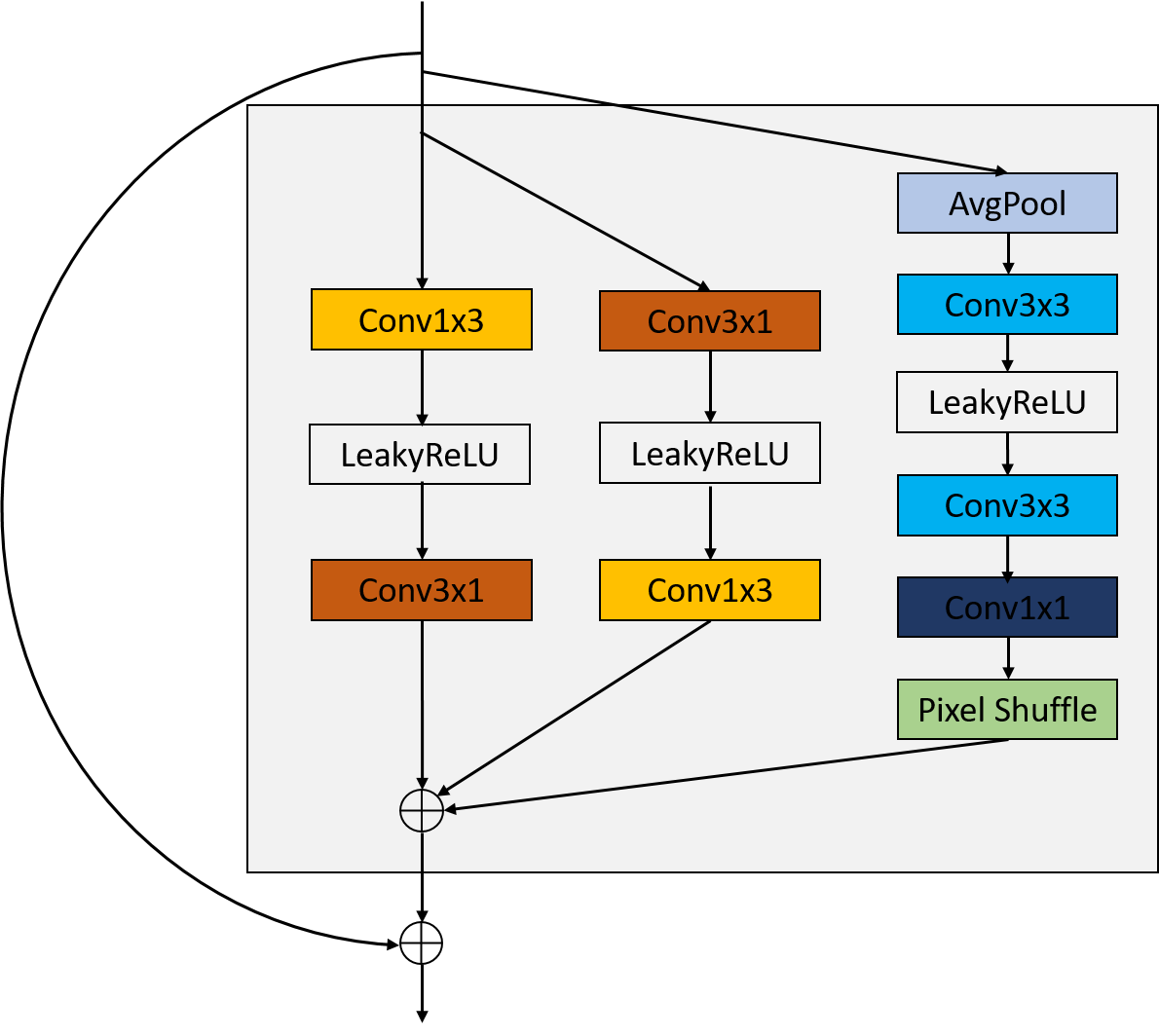}}
\subfigure[]{\label{fig:fig:fcp}\includegraphics[width=.66\linewidth]{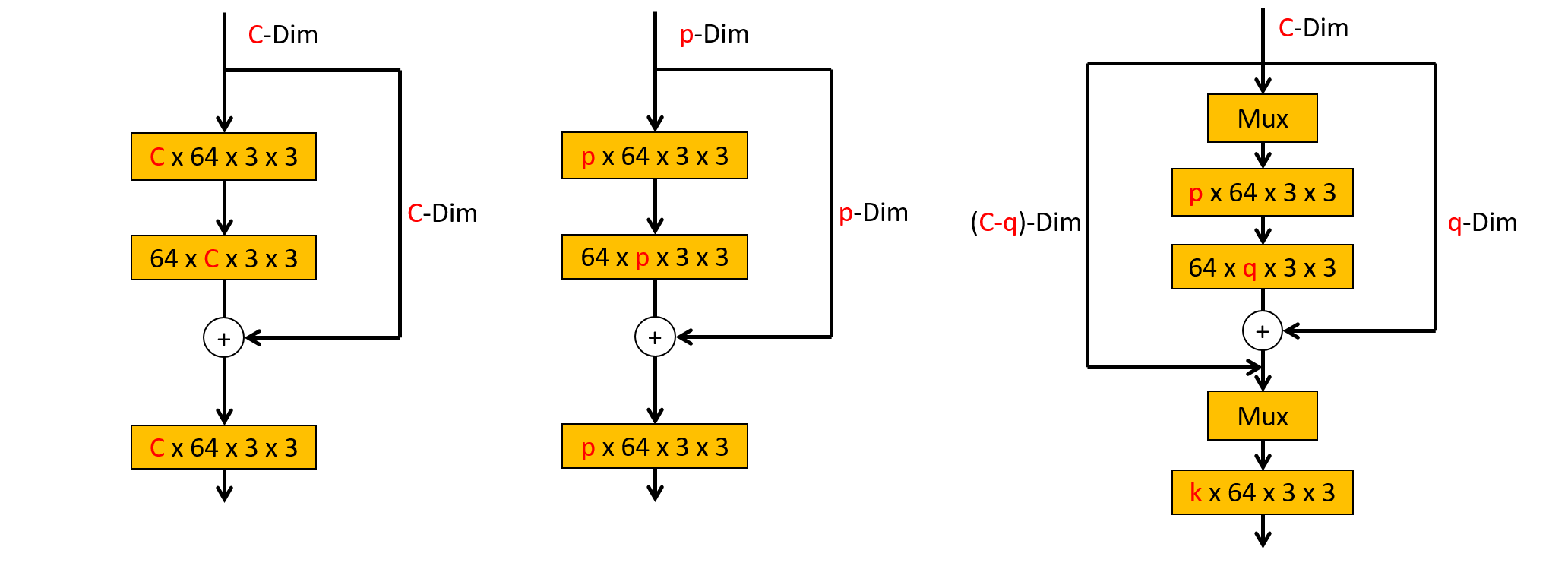}}
\caption{SC-CVLAB Team: (a) the architecture of  hybrid composition; (b) an illustration of baseline structure, structure pruned by group strategy, and structure pruned by the proposed fine-grained strategy.}
\label{fig:HCSRN_arch1}
\end{figure}

\begin{figure}[!ht]
\centering
\includegraphics[width=.7\linewidth]{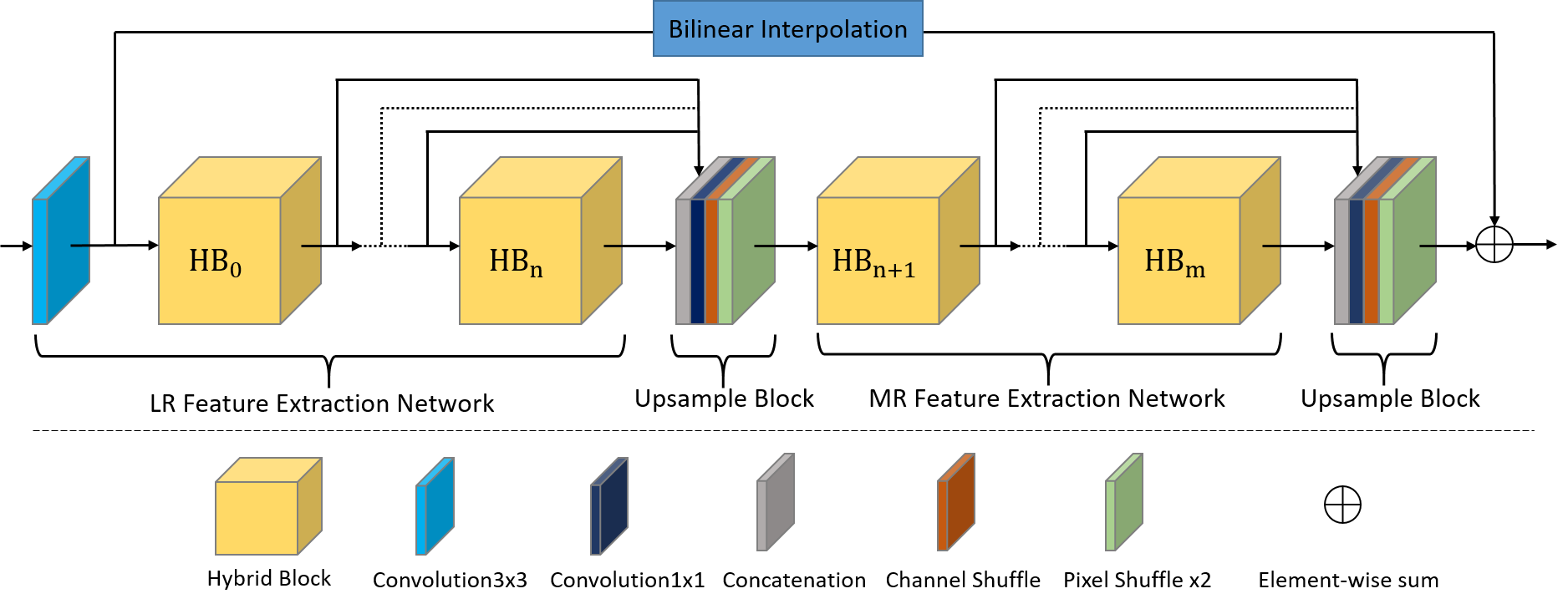}
\caption{SC-CVLAB Team: the network architecture.}
\label{fig:HCSRN_arch}
\end{figure}

The SC-CVLAB team proposed \textbf{Adaptive Hybrid Composition Based Super-Resolution Network via Fine-grained Channel Pruning}~\cite{chen2020adaptive}. Firstly, a hybrid composition based SR neural network (HCSRN) is designed. As shown in Fig.~\ref{fig:HB_arch}, the hybrid composition is constructed by three parallel blocks. The element-wise sum operation is conducted for local feature fusion. The whole block adopts residual connection. The asymmetric kernels are adopted. Instead of using $5\times1$ and $1\times5$ kernels, kernels with smaller sizes are used to factorize the normal $3\times3$ convolution into an $3\times1$ and a $1\times3$ convolution. To extract different features without significant performance drop, two asymmetric blocks with inverted kernel order is utilized. Another way to reduce model size is reducing the scale of feature. Thus an average pooling followed by ``Conv-LeakyReLU-Conv'' is utilized, and sub-pixel convolution is used to reconstruct the HR image. Then the proposed HCSRN is constructed based on this hybrid module with progressive upsampling strategy as shown in Fig.~\ref{fig:HCSRN_arch}. 

Secondly, the over-parameterized HCSRN is used as the baseline model and channel pruning is applied to it to further reduce model size. 
The pruning criterion in~\cite{you2019gate} is utilized. 
Instead of employing the group pruning strategy~\cite{ding2019centripetal,li2020group}, the gating function is asserted before and after each convolution. Thus each channel is allowed to be pruned independently. To avoid the misalignment problem between the convolution and the skip connection, the skip connections will not be pruned and the residual addition operation should always be kept. 
The difference between the proposed fine-grained channel pruning strategy and the previous grouping method is shown in Fig.~\ref{fig:fig:fcp}.
Finally, a pruned lightweight model called adaptive hybrid composition based super-resolution network (AHCSRN) is obtained.

\subsection*{HiImageTeam}
\begin{figure}[!ht]
\centering
\subfigure[]{\label{fig:HiImageTeam}\includegraphics[width=.5\linewidth]{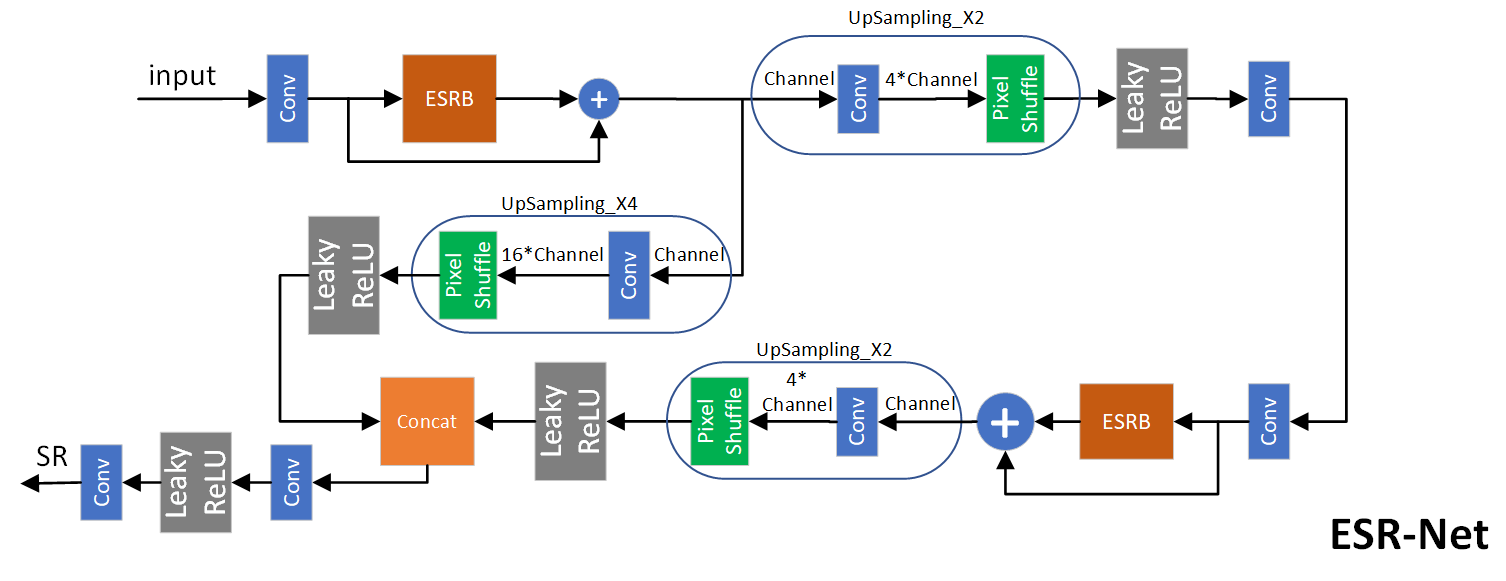}}
\subfigure[]{\label{fig:HiImageTeam_fig2}\includegraphics[width=.48\linewidth]{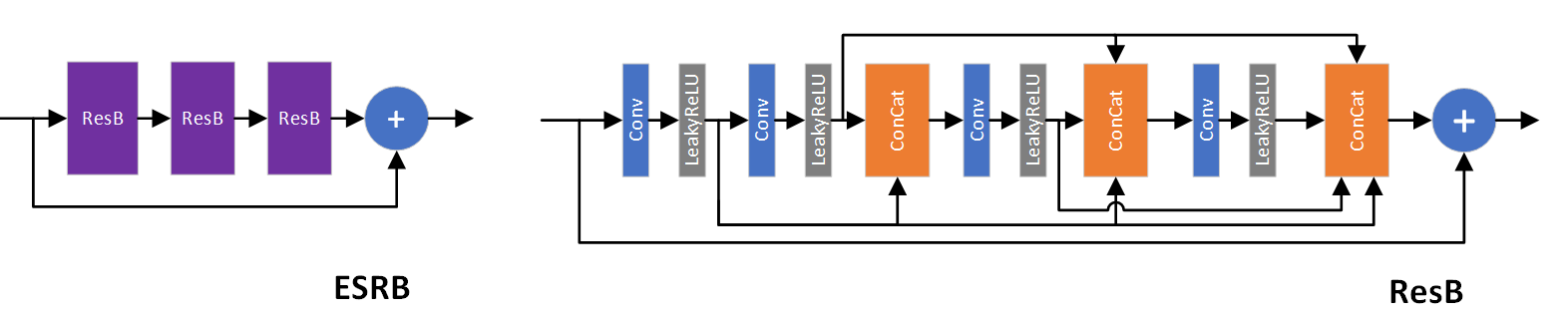}}
\caption{HiImageTeam Team: (a) the network architecture; (b) the ESR Block and Residual Block.}
\end{figure}

The HiImageTeam team proposed \textbf{Efficient SR-Net (ESR-Net)} for the challenge. As shown in Fig.~\ref{fig:HiImageTeam}, ESR-Net achieves an upscaling factor of 4 via two successive $\times2$ subnetworks (ESRB). In order to improve the performance, residual dense network shown in Fig.~\ref{fig:HiImageTeam_fig2} is adopted. L1 loss is used to train ESR-Net. 

\subsection*{SAMSUNG\_TOR\_AIC}
\begin{figure}[!ht]
\centering
\includegraphics[width=0.8\linewidth]{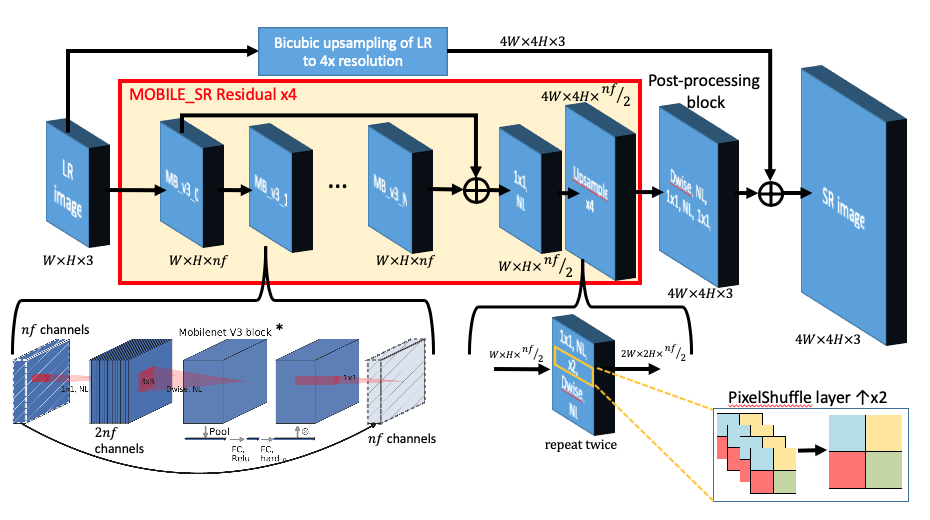}
\caption{SAMSUNG\_TOR\_AIC Team: the network architecture.}
\label{fig:method}
\end{figure}

The SAMSUNG\_TOR\_AIC team proposed \textbf{SAM\_SR\_LITE}, a Lightweight MobileNetV3 network for Efficient Super-Resolution~\cite{wang2020efficientSR}.
The core of this approach is to use modified MobileNetV3~\cite{howard2019searching} blocks to design an efficient method for SR. The authors found that for the MobileNetV3 architecture batch normalization layers improved the performance. The architecture takes the LR images as input and consists of N+1 modified MobileNetV3 blocks, with a skip connection (addition) from the output of the first block to the output of the last block. The output of the last block is then upscaled using the two PixelShuffle operations \cite{shi2016real} with non-linearities and convolutions. To yield three color channels, a post-processing block containing a depthwise separable convolution and a $1\times 1$ convolution layer is applied to the output of the upscaled result. Finally, there is an additional skip connection between the bicubic upsampling of the LR input and the output of the post processing block. Fig.~\ref{fig:method} provides a overview of the architecture.

The hyperparameters include the number of blocks $N$, the number of feature mapss $n_f$ in each block, and the expansion factor in each block. For the final submission, $N=16$, $n_f=72$, and the block expansion factor is $2$.

\subsection*{neptuneai}
The model is based on the MSRResNet and the basic blocks in MSRResNet is changed to Inverted Residuals Block in MobileNetV2. This architecture is used as backbone network. Then NAS is used to discover the best architecture for lightweight SR tasks.
A search space which is composed of several kinds of blocks is designed. 
The basic blocks for NAS include
\begin{enumerate}
    \item Inverted Residuals Block with 3 expand ration,
    \item Inverted Residuals Block with 6 expand ration,
    \item Basic Residual Block,
    \item Basic Residual Block with leaky ReLU activation function.
\end{enumerate}

\subsection*{lyl}
The lyl team proposed \textbf{Coarse to Fine network (CFN)} progressive super-resolution reconstruction, which contains a CoarseNet and a FineNet. The FineNet contains a lightweight upsampling  module (LUM).
There are two characteristics in CFN, \ie the progressiveness and the merging of the output of the LUM to correct the input in each level. 
Such progressive cause-and-effect process helps to achieve the principle for image SR. That is, high-level information can guide an LR image to recover a better SR image. 
In the proposed network, there are three indispensable parts: 1) tying the loss at each level, 2) using LUM and 3) providing feature maps extracted in lower level to ensure the availability of low-level information.

\subsection*{CET\_CVLab}
\begin{figure}[!ht]
   \centering
   \includegraphics[width=0.7\textwidth]{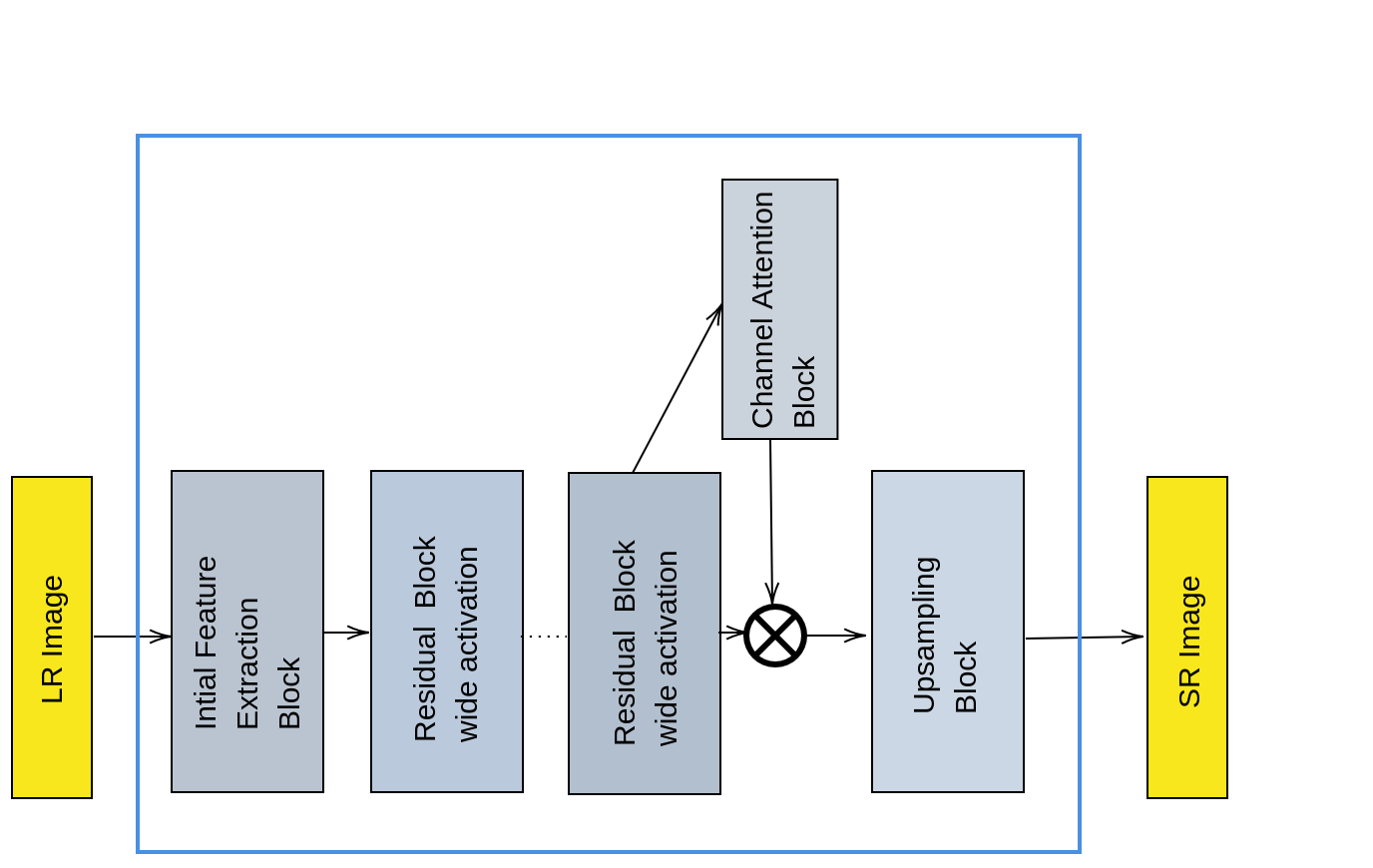}
   \caption{CET\_CVLab Team: the network architecture.}
   \label{fig:cet}
\end{figure}

The architecture used is inspired by wide activation based network and channel attention network. The network, as shown in Fig.~\ref{fig:cet},  mainly consists of 3 blocks, a feature extraction block, a series of wide activation residual blocks and a set of progressive upsampling blocks ($\times2$). The expansion factor used for wide activation block is six. The depth within the feature extraction blocks and wide activation blocks is 32. The network contains 1.378 million trainable parameters. Charbonnier loss is used for training the network as it captures the edge information better than the mean squared error (MSE) loss.

\subsection*{InnoPeak\_SR}
\begin{figure}[!ht]
   \centering
   \includegraphics[width=0.8\textwidth]{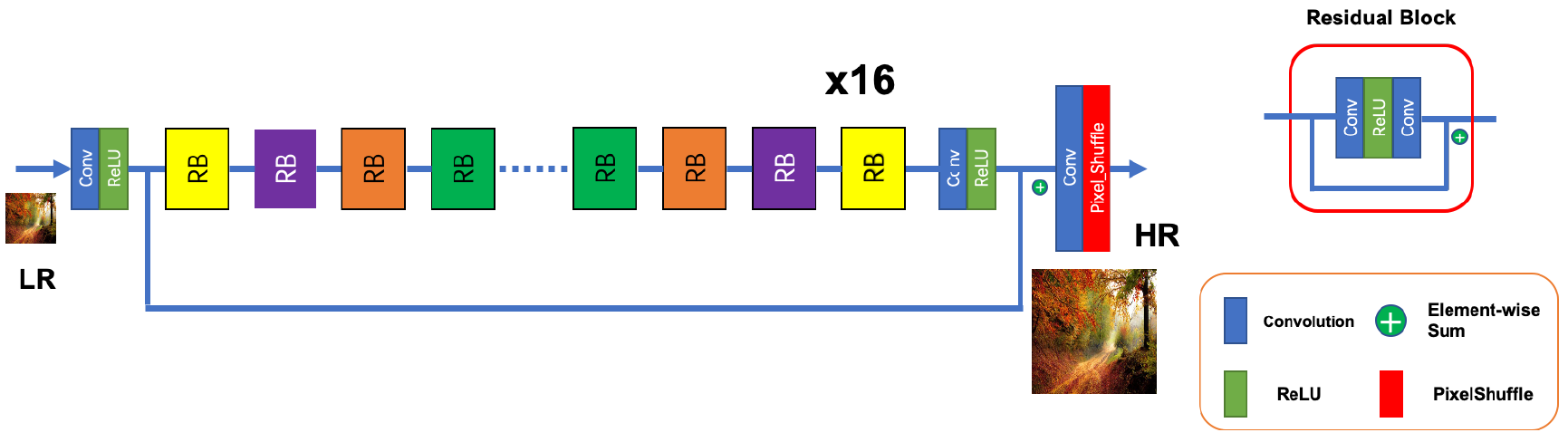}
   \caption{InnoPeak\_SR Team: an illustration of the proposed SRRN. SRRN consists of 16 residual blocks, but only have four different weights. The residual blocks with same color share the same parameter weights.}
   \label{fig:G}
\end{figure}

The InnoPeak\_SR team proposed Shuffled Recursive Residual Network (SRRN) for Efficient Image Super-Resolution. As shown in Fig.~\ref{fig:G}, the proposed SRRN mainly consists of four parts: shallow feature extraction, residual block feature extraction, skip connection module, and upscale reconstruction module.
To extract the shallow features, one $3\times3$ convolution layer (pad 1, stride 1, and channel 64) is used, which is followed by one Leaky ReLU layer (slop 0.1). To extract the mid-level and high-level features, 16 weights shared recursive residual blocks are used.
The residual block consists of two $3\times3$ convolution layer (pad 1, stride 1, and channel 64). Only the first convolution layer is followed by a Leaky ReLU layer (slop 0.1).
Batch Normalization (BN) layers are not used.
In this work, one $3\times3$ convolution layer (pad 1, stride 1, and channel 64) and one sub-pixel convolution layer is used as upscaling module and reconstruction module.
During the training, HR patches of size $196\times196$ are randomly cropped from HR images, and the mini-batch size is set to 25. The proposed model is trained by minimizing L1 loss function with Adam optimizer. 

\subsection*{Summer}
\begin{figure}[!ht]
  \centering
  \includegraphics[width=0.8\textwidth]{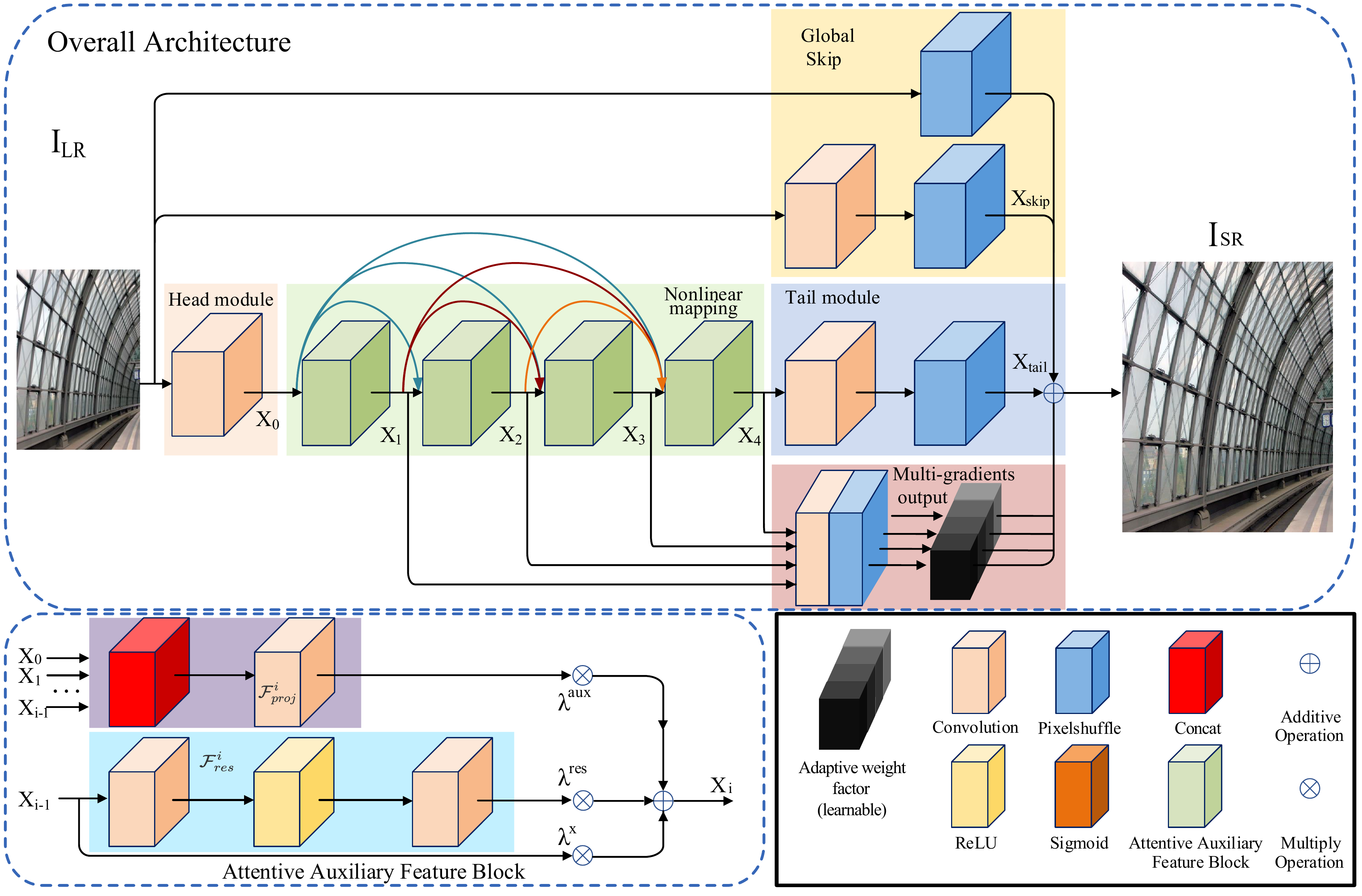}
  \caption{Summer Team: the network architecture.}
  \label{netArch}
\end{figure}

The Summer team proposed \textbf{AMAF-Net}. 
The model consists of four parts: 1) Attentive Auxiliary Feature Block (AAF Block) reuses all features of the preceding blocks as the input feature map of the current layer. By applying convolution to the concatenated feature maps, the network learns how to combine the information from different blocks. 2) Global residual connection. The input image are directly upsampled by PixelShuffler, which can be regarded as a part of the output image. It brings the most basic information to the output. 3) Multi-gradients skip connection. The outputs of each block are upsampled by $1\times1$ convolution and PixelShuffler. The results are also a proportion of the output, which are useful for transmitting different frequency information. Meanwhile, gradients can be propagated from the tail of the network to the head. 4) By using the adaptive weight factor multiple outputs are combined with the learnable parameters, which can adaptively determine the contributions of each blocks. The network architecture is shown in Fig.~\ref{netArch}.

L1 loss is used in the training process. 
To further reduce parameters and FLOPs, the convolution of last two blocks is replaced by $1\times1$ convolution.

\subsection*{Zhang9678}
The Zhang9678 team proposed \textbf{LFN} which contains three parts, including shallow feature extraction, deep feature fusion and upsampling. Shallow feature extraction contains a convolutional layer, which is used to map images from image space to feature space. Deep feature fusion contains four DB blocks and one modified convLSTM. The feature map obtained by the shallow feature extraction of the image passes through four consecutive DB modules to obtain feature maps in different depths.
LSTM is usually used to process time series data. For time series data, there is a high degree of similarity between the time steps, showing a progressive relationship on the time dimension. The feature maps at different levels are regarded as a kind of pseudo-time series data because they also contain a progressive relationship at different levels and their similarity is also high.
Specifically, the hidden state $h$ and cell state $c$ of the initial LSTM are initialized to 0. For each state update, a feature map is selected and sent to the LSTM along with the corresponding $h$ and $c$. After the operations in the LSTM, the new $h$ and $c$ are obtained. This cycle is repeated for four times in total. 
The resulting $h$ is concatenated with the shallow features. Then a $1\times1$ convolution and a $3\times3$ convolution are appended to get the feature map with 48 channels. Finally, a PixelShuffler is used to get the final SR image.

\begin{figure}[!ht]
    \centering
    \includegraphics[width=0.7\textwidth]{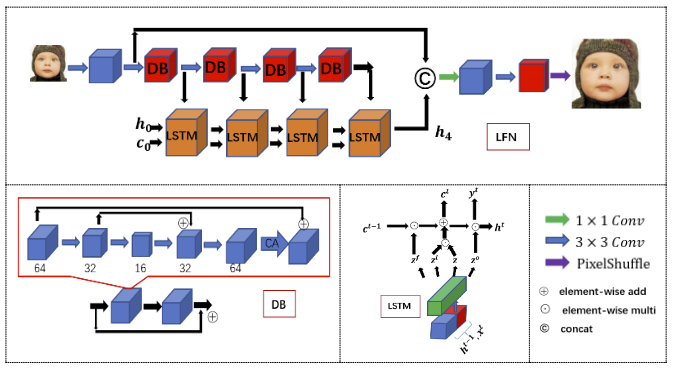}
    \caption{Zhang9678 Team: the network architecture.}
    \label{fig:2}
\end{figure}

\subsection*{H-ZnCa}
\begin{figure}[!ht]
	\begin{center}
\subfigure[]{\label{structure}\includegraphics[width=.65\linewidth]{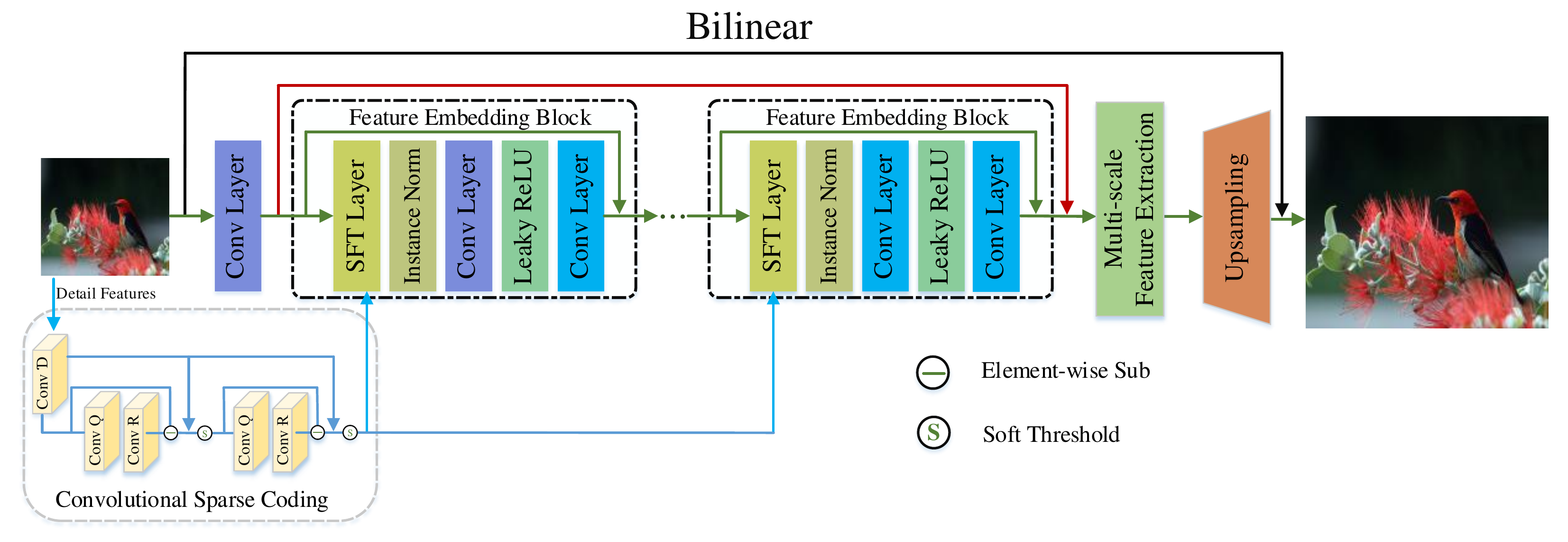}}
\subfigure[]{\label{MFEM}\includegraphics[width=.3\linewidth]{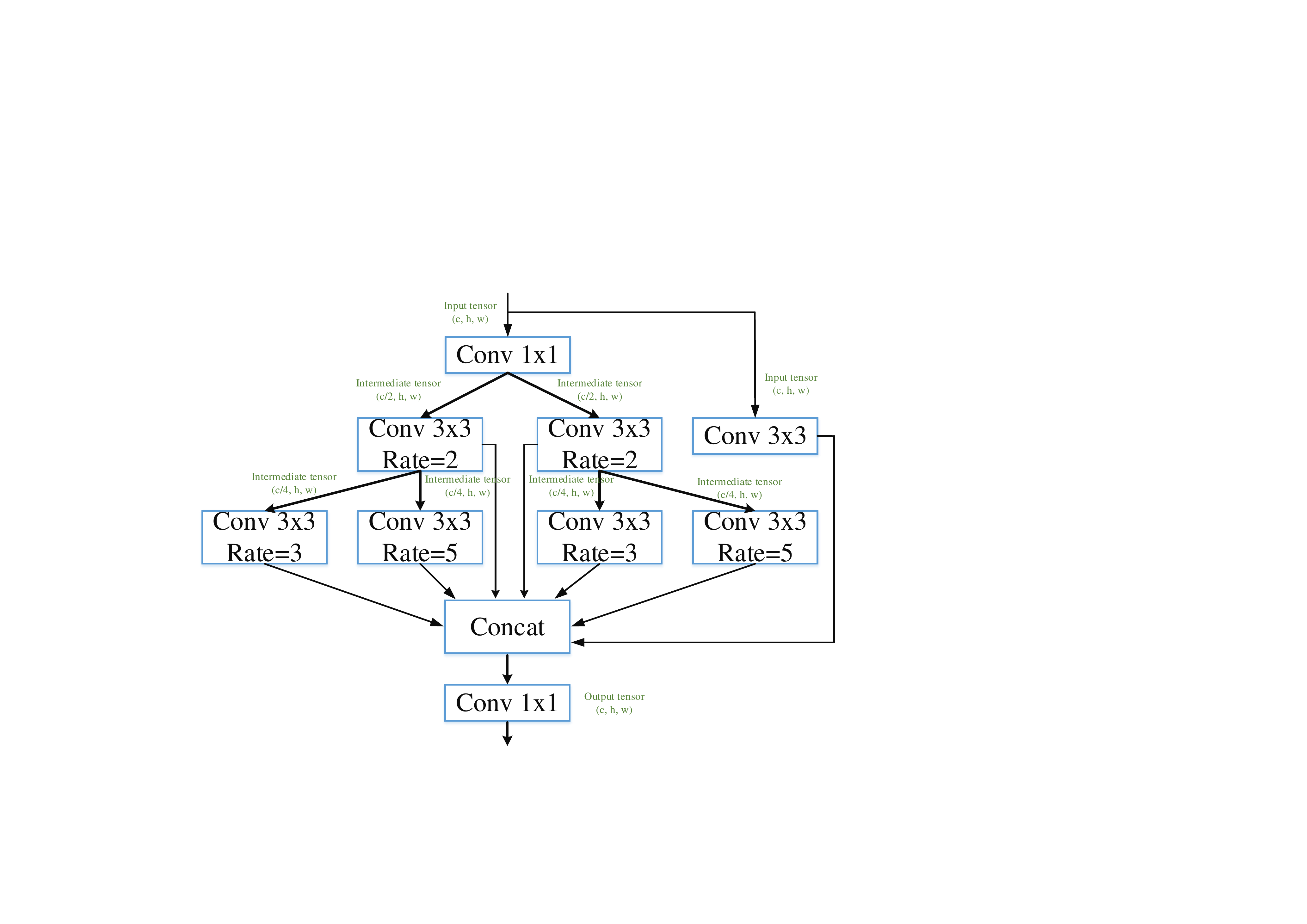}}
 		\caption{H-ZnCa Team: (a) the architecture of SPSR Network; (b) the multi-scale feature extraction module (MFEM).}
	\end{center}
\end{figure}

The H-ZnCa team proposed \textbf{Sparse Prior-based Network for Efficient Image Super-Resolution}. As shown in Fig.~\ref{structure}, the proposed lightweight model, named SPSR, consists of three components: high-frequency sparse coding generation, feature embedding, and multi-scale feature extraction. Specifically, a convolutional sparse coding module (CSCM)~\cite{csc,rcsc} is first performed to obtain the high-frequency spare representation of the input. Then, a feature embedding block (FEB)~\cite{wang2018recovering} which consists of a Spatial Feature Transform (SFT) layer and two convolutional layers is designed for spatial-wise feature modulation conditioned on the sparse prior representation. To further enhance the abstract ability of SPSR, a multi-scale feature extraction module (MFEM) with channel split mechanism is proposed to efficiently utilize the hierarchical features. As shown in Fig.~\ref{MFEM}, MFEM contains several convolutions with different dilation factors.

\subsection*{MLP\_SR}
\begin{figure}[!ht]
  \centering
  \includegraphics[width=0.8\textwidth]{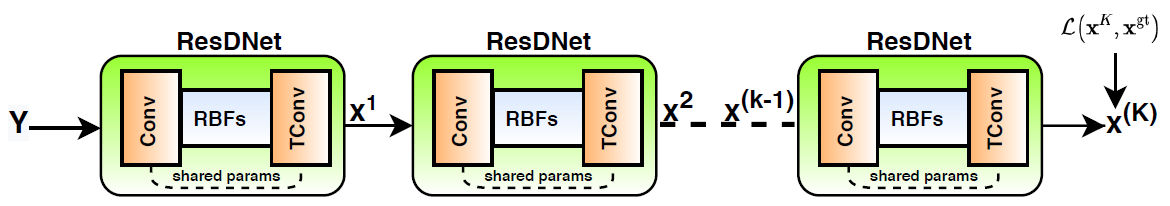}
  \caption{MLP\_SR Team: the architecture of the iterative SR approach ISRResDNet.}
  \label{mlpsr}
\end{figure}

The MLP\_SR team proposed a lightweight deep iterative SR learning method
(ISRResDNet) that solves the SR task as a sub-solver of image denoising by
the residual denoiser networks~\cite{lefkimmiatis2018universal}. It is inspired by powerful image regularization and large-scale optimization techniques used to solve general inverse problems. 
The proposed iterative SR approach is shown in Fig.~\ref{mlpsr}. The authors unroll the ResDNet~\cite{lefkimmiatis2018universal} into $K$ stages and each stage performs the PGM updates.

\section*{Acknowledgements}
We thank the AIM 2020 sponsors: 
HUAWEI, MediaTek, Google, NVIDIA, Qualcomm, and Computer Vision Lab (CVL) ETH Zurich.

\appendix
\section{Teams and affiliations}
\label{sec:teams}

\subsection*{AIM2020 team}
\noindent{\textbf{Title: }} AIM 2020 Efficient Super-Resolution Challenge\\
\noindent{\textbf{Members: }} \\
Kai Zhang (\href{mailto:kai.zhang@vision.ee.ethz.ch}{kai.zhang@vision.ee.ethz.ch}),\\
Martin Danelljan (\href{mailto:martin.danelljan@vision.ee.ethz.ch}{martin.danelljan@vision.ee.ethz.ch}),\\
Yawei Li (\href{mailto:yawei.li@vision.ee.ethz.ch}{yawei.li@vision.ee.ethz.ch}),\\
Radu Timofte (\href{mailto:radu.timofte@vision.ee.ethz.ch}{radu.timofte@vision.ee.ethz.ch})\\
\noindent{\textbf{Affiliations: }}\\
Computer Vision Lab, ETH Zurich, Switzerland\\

\subsection*{NJU\_MCG}
\noindent{\textbf{Title: }}Residual Feature Distillation Network (RFDN) \\
\noindent{\textbf{Members: }}{Jie Liu
\\\noindent(\href{mailto:username@mail.com}{jieliu@smail.nju.edu.cn})}, Jie Tang, Gangshan Wu \\
\noindent{\textbf{Affiliation: }}\\
State Key Laboratory for Novel Software Technology, Nanjing University, Nanjing 210023, China\\

\subsection*{AiRiA\_CG}
\noindent{\textbf{Title: }}Faster Information Multi-distillation Network via Asymmetric Convolution\\
\noindent{\textbf{Members: }} Yu Zhu\\\noindent(\href{mailto:zhuyu.cv@gmail.com}{zhuyu.cv@gmail.com}), Xiangyu He, Wenjie Xu, Chenghua Li, Cong Leng, Jian Cheng\\
\noindent{\textbf{Affiliation: }}\\
Nanjing Artificial Intelligence Chip Research, Institute of Automation, Chinese Academy of Sciences (AiRiA); MAICRO\\

\subsection*{UESTC-MediaLab}
\noindent{\textbf{Title: }}Efficient Super-Resolution with Gradually Kernel Dilution\\
\noindent{\textbf{Members: }}{Guangyang Wu$^{1}$
\\\noindent(\href{mailto:username@mail.com}{mulns@outlook.com})}, Wenyi Wang$^{1}$, Xiaohong Liu$^{2}$\\
\noindent{\textbf{Affiliation: }}\\
$^{1}$ University of Electronic Science and Technology of China\\
$^{2}$ McMaster University
\\

\subsection*{XPixel}
\noindent{\textbf{Title: }}Efficient Image Super-Resolution using Pixel Attention
\\
\noindent{\textbf{Members: }}{Hengyuan Zhao
\\\noindent(\href{mailto:hy.zhao1@siat.ac.cn}{hy.zhao1@siat.ac.cn})}, 
Xiangtao Kong, Jingwen He,Yu Qiao, Chao Dong\\
\noindent{\textbf{Affiliation: }}\\
Shenzhen Institutes of Advanced Technology, Chinese Academy of Sciences
\\

\subsection*{HaiYun}
\noindent{\textbf{Title: }}Lightweight Image Super-resolution with Lattice Block\\
\noindent{\textbf{Members: }}{Xiaotong Luo
\\\noindent(\href{mailto:xiaotluo@qq.com}{xiaotluo@qq.com})}, 
Liang Chen, Jiangtao Zhang\\
\noindent{\textbf{Affiliation: }}\\
Xiamen University, China
\\

\subsection*{IPCV\_IITM}
\noindent{\textbf{Title: }}Lightweight Attentive Residual Network for Image Super-Resolution
\\
\noindent{\textbf{Members: }}{Maitreya Suin
\\\noindent(\href{mailto:maitreyasuin21@gmail.com}{maitreyasuin21@gmail.com})}, 
Kuldeep Purohit, A. N.
Rajagopalan\\
\noindent{\textbf{Affiliation: }}\\
Indian Institute of Technology Madras, India\\

\subsection*{404NotFound}
\noindent{\textbf{Title: }}GCSR
\\
\noindent{\textbf{Members: }}{Xiaochuan Li
\noindent(\href{mailto:1182784700@qq.com}{1182784700@qq.com})}\\
\noindent{\textbf{Affiliation: }}\\
Nanjing University of Aeronautics and Astronautics,
Nanjing, China\\

\subsection*{MDISL-lab}
\noindent{\textbf{Title: }}PFSNet: Partial Features Sharing for More Efficient Super-Resolution\\
\noindent{\textbf{Members: }}{Zhiqiang Lang
\\\noindent(\href{mailto:username@mail.com}{2015303107lang@mail.nwpu.edu.cn})}, Jiangtao Nie, Wei Wei, Lei Zhang\\
\noindent{\textbf{Affiliation: }}\\
School of Computer Science, Northwestern Polytechnical University, China\\
\\

\subsection*{MLVC}
\noindent{\textbf{Title: }}Multi Attention Feature Fusion Super-Resolution Network\\
\noindent{\textbf{Members: }}{Abdul Muqeet$^{1}$
\\\noindent(\href{mailto:amuqeet@khu.ac.kr}{amuqeet@khu.ac.kr})}, Jiwon Hwang$^{1}$, Subin Yang$^{1}$, JungHeum Kang$^{1}$, Sungho Bae$^{1}$, Yongwoo Kim$^{2}$\\
\noindent{\textbf{Affiliation: }}\\
$^{1}$ Kyung Hee University, Republic of Korea\\
$^{2}$ Sang Myung University, Republic of Korea\\

\subsection*{XMUlab}
\noindent{\textbf{Title: }}Pixelshuffle Attention Network\\
\noindent{\textbf{Members: }}{Liang Chen
\\\noindent(\href{1806668306@qq.com}{1806668306@qq.com})}, Jiangtao Zhang, Xiaotong Luo, Yanyun Qu\\
\noindent{\textbf{Affiliation: }}\\
Xianmen University\\
\\

\subsection*{MCML-Yonsei}
\noindent{\textbf{Title: }}LarvaNet: Hierarchical Super-Resolution via Internal Output and Loss\\
\noindent{\textbf{Members: }}{Geun-Woo Jeon
\\\noindent(\href{mailto:geun-woo.jeon@yonsei.ac.kr}{geun-woo.jeon@yonsei.ac.kr})}, Jun-Ho Choi, Jun-Hyuk Kim, Jong-Seok Lee\\
\noindent{\textbf{Affiliation: }}\\
Yonsei University, Republic of Korea\\
\\

\subsection*{LMSR}
\noindent{\textbf{Title: }}LMSR\\
\noindent{\textbf{Members: }}{Steven Marty \\\noindent(\href{mailto:martyste@student.ethz.ch}{martyste@student.ethz.ch})}, Eric Marty\\
\noindent{\textbf{Affiliation: }}\\
ETH Zurich

\subsection*{ZJUESR2020}
\noindent{\textbf{Title: }}IdleSR: Efficient Super-Resolution Network with Multi-Scale IdleBlocks\\
\noindent{\textbf{Members: }} Dongliang Xiong
(\href{mailto:xiongdl@zju.edu.cn}{xiongdl@zju.edu.cn})\\
\noindent{\textbf{Affiliation: }}
Zhejiang University\\
\\

\subsection*{SC-CVLAB}
\noindent{\textbf{Title: }}Adaptive Hybrid Composition Based Super-Resolution Network via Fine-grained Channel Pruning\\
\noindent{\textbf{Members: }}{Siang Chen
\noindent(\href{mailto:11631032@zju.edu.cn}{11631032@zju.edu.cn})}\\
\noindent{\textbf{Affiliation: }}
Zhejiang University\\

\subsection*{HiImageTeam}
\noindent{\textbf{Title: }}Efficient SR-Net\\
\noindent{\textbf{Members: }}{Lin Zha$^{1}$
\\\noindent(\href{mailto:zhalin@hisense.com}{zhalin@hisense.com})}, Jiande Jiang$^{1}$, Xinbo Gao$^{2}$, Wen Lu$^{2}$\\
\noindent{\textbf{Affiliation: }}\\
$^{1}$ Qingdao Hi-image Technologies Co.,Ltd (Hisense Visual Technology Co.,Ltd.)\\
$^{2}$ Xidian University
\\

\subsection*{SAMSUNG\_TOR\_AIC}
\noindent{\textbf{Title: }}Lightweight MobileNetV3 for Efficient Super-Resolution\\
\noindent{\textbf{Members: }}{Haicheng Wang
\\\noindent(\href{mailto:h.wang1@samsung.com}{h.wang1@samsung.com})}, Vineeth Bhaskara, Alex Levinshtein, Stavros Tsogkas, Allan Jepson\\
\noindent{\textbf{Affiliation: }}
Samsung AI Centre, Toronto
\\

\subsection*{neptuneai}
\noindent{\textbf{Title: }}Lightweight super resolution network with Neural Architecture Search\\
\noindent{\textbf{Members: }}{Xiangzhen Kong
\noindent(\href{mailto:neptune.team.ai@gmail.com}{neptune.team.ai@gmail.com})}
\\

\subsection*{lyl}
\noindent{\textbf{Title: }}Coarse to Fine Pyramid Networks for Progressive Image Super-Resolution\\
\noindent{\textbf{Members: }}{Tongtong Zhao$^{1}$
\\\noindent(\href{mailto:daitoutiere@gamil.com}{yaopuss@126.com})}, Shanshan Zhao$^{2}$\\
\noindent{\textbf{Affiliation: }}\\
$^{1}$ Dalian Maritime University\\
$^{2}$ China Everbright Bank Co., Ltd
\\

\subsection*{CET\_CVLab}
\noindent{\textbf{Title: }}Efficient Single Image Super-resolution using Progressive Wide
Activation Net\\
\noindent{\textbf{Members: }}{Hrishikesh P S \\\noindent(\href{mailto:hrishikeshps94@gmail.com}{hrishikeshps94@gmail.com})}, Densen Puthussery, Jiji C V\\
\noindent{\textbf{Affiliation: }}\\
College of Engineering, Trivandrum\\

\subsection*{wozhu}
\noindent{\textbf{Title: }}FSSR\\
\noindent{\textbf{Members: }}{Nan Nan \\\noindent(\href{mailto:2829272117@qq.com}{2829272117@qq.com})}, Shuai Liu\\

\subsection*{InnoPeak\_SR}
\noindent{\textbf{Title: }} Shuffled Recursive Residual Network for Efficient Image Super-Resolution\\
\noindent{\textbf{Members: }}{Jie Cai
\\\noindent(\href{mailto:caijie0620@gmail.com}{caijie0620@mail.com})}, Zibo Meng, Jiaming Ding, Chiu Man Ho\\
\noindent{\textbf{Affiliation: }}\\
InnoPeak Technology, Inc.\\

\subsection*{Summer}
\noindent{\textbf{Title: }}Adaptively Multi-gradients Auxiliary Feature Learning for Efficient Super-resolution\\
\noindent{\textbf{Members: }}{Xuehui Wang$^{1,2}$
\\\noindent(\href{mailto:wangxh228@mail2.sysu.edu.cn}{wangxh228@mail2.sysu.edu.cn})}, Qiong Yan$^{1}$, Yuzhi Zhao$^{3}$, Long Chen$^{2}$\\
\noindent{\textbf{Affiliation: }}\\
$^{1}$ SenseTime Research\\
$^{2}$ Sun Yat-sen University\\
$^{3}$ City University of Hong Kong
\\

\subsection*{Zhang9678}
\noindent{\textbf{Title: }}Lightweight super-resolution network using convLSTM fusion features\\
\noindent{\textbf{Members: }}{Jiangtao Zhang
\\\noindent(\href{mailto:username@mail.com}{1328937778@qq.com})},\\Xiaotong Luo, Liang Chen, Yanyun Qu\\
\noindent{\textbf{Affiliation: }}\\
Xianmen University\\

\subsection*{H-ZnCa}
\noindent{\textbf{Title: }}Sparse Prior-based Network for Efficient Image
Super-Resolution\\
\noindent{\textbf{Members: }}{Long Sun
\\\noindent(\href{lungsuen@163.com}{lungsuen@163.com})}, Wenhao Wang,  Zhenbing Liu, Rushi Lan\\
\noindent{\textbf{Affiliation: }}\\
Guilin University of Electronic Technology, Guilin 541004, China.\\
\\

\subsection*{MLP\_SR}
\noindent{\textbf{Title: }}A Light-weight Deep Iterative Residual Convolutional Network for Super-Resolution\\
\noindent{\textbf{Members: }}{Rao Muhammad Umer
\\\noindent(\href{mailto:engr.raoumer943@gmail.com}{engr.raoumer943@gmail.com})}, Christian Micheloni\\
\noindent{\textbf{Affiliation: }}\\
University of Udine, Italy\\

\bibliographystyle{splncs04}
\bibliography{egbib}
\end{document}